\documentclass[aps,prb, twocolumn,superscriptaddress]{revtex4-2}
\usepackage{graphicx}
\usepackage[hidelinks]{hyperref} 
\usepackage{bm}
\usepackage{amsmath}

\graphicspath{{/Users/vig/Documents/Paper/201028_CuInCr4S8_lowE/plots/}}

\begin{document}


\title{Hierarchical excitations from correlated spin tetrahedra on the breathing pyrochlore lattice} 
\thanks{This manuscript has been authored by UT-Battelle, LLC under Contract No. DE-AC05-00OR22725 with the U.S. Department of Energy.  The United States Government retains and the publisher, by accepting the article for publication, acknowledges that the United States Government retains a non-exclusive, paid-up, irrevocable, world-wide license to publish or reproduce the published form of this manuscript, or allow others to do so, for United States Government purposes.  The Department of Energy will provide public access to these results of federally sponsored research in accordance with the DOE Public Access Plan (http://energy.gov/downloads/doe-public-access-plan).}%

\renewcommand*{\thefootnote}{\arabic{footnote}}

\author{Shang Gao}
\affiliation{Materials Science \& Technology Division, Oak Ridge National Laboratory, Oak Ridge, TN 37831, USA}
\affiliation{Neutron Scattering Division, Oak Ridge National Laboratory, Oak Ridge, TN 37831, USA}

\author{Andrew F. May}
\affiliation{Materials Science \& Technology Division, Oak Ridge National Laboratory, Oak Ridge, TN 37831, USA}

\author{Mao-Hua Du}
\affiliation{Materials Science \& Technology Division, Oak Ridge National Laboratory, Oak Ridge, TN 37831, USA}

\author{Joseph A. M. Paddison}
\affiliation{Materials Science \& Technology Division, Oak Ridge National Laboratory, Oak Ridge, TN 37831, USA}

\author{Hasitha Suriya Arachchige}
\affiliation{Department of Physics \& Astronomy, University of Tennessee, Knoxville, TN 37996, USA}
\affiliation{Materials Science \& Technology Division, Oak Ridge National Laboratory, Oak Ridge, TN 37831, USA}

\author{Ganesh Pokharel}
\affiliation{Department of Physics \& Astronomy, University of Tennessee, Knoxville, TN 37996, USA}
\affiliation{Materials Science \& Technology Division, Oak Ridge National Laboratory, Oak Ridge, TN 37831, USA}

\author{Clarina dela Cruz}
\affiliation{Neutron Scattering Division, Oak Ridge National Laboratory, Oak Ridge, TN 37831, USA}

\author{Qiang Zhang}
\affiliation{Neutron Scattering Division, Oak Ridge National Laboratory, Oak Ridge, TN 37831, USA}

\author{Georg Ehlers}
\affiliation{Neutron Technologies Division, Oak Ridge National Laboratory, Oak Ridge, TN 37831, USA}

\author{David S. Parker}
\affiliation{Materials Science \& Technology Division, Oak Ridge National Laboratory, Oak Ridge, TN 37831, USA}

\author{David G. Mandrus}
\affiliation{Department of Physics \& Astronomy, University of Tennessee, Knoxville, TN 37996, USA}
\affiliation{Materials Science \& Technology Division, Oak Ridge National Laboratory, Oak Ridge, TN 37831, USA}
\affiliation{Department of Material Science \& Engineering, University of Tennessee, Knoxville, TN 37996, USA}

\author{Matthew B. Stone}
\affiliation{Neutron Scattering Division, Oak Ridge National Laboratory, Oak Ridge, TN 37831, USA}

\author{Andrew D. Christianson}
\affiliation{Materials Science \& Technology Division, Oak Ridge National Laboratory, Oak Ridge, TN 37831, USA}


\date{\today}

\pacs{}

\begin{abstract}

The hierarchy of the coupling strengths in a physical system often engenders an effective model at low energies where the decoupled high-energy modes are integrated out. Here, using neutron scattering, we show that the spin excitations in the breathing pyrochlore lattice compound CuInCr$_4$S$_8$ are hierarchical and can be approximated by an effective model of correlated tetrahedra at low energies. At higher energies, intra-tetrahedron excitations together with strong magnon-phonon couplings are observed, which suggests the possible role of the lattice degree of freedom in stabilizing the spin tetrahedra. Our work illustrates the spin dynamics in CuInCr$_4$S$_8$ and demonstrates a general effective-cluster approach to understand the dynamics on the breathing-type lattices. 

\end{abstract}

\maketitle

For the description of a physical system, selecting an appropriate energy scale is always the first step as it determines what (quasi-)particles and interactions might play a role. A well-known example is the Standard Model, where massive elementary particles emerge at `low' energies through spontaneous symmetry breaking. In solids, pertinent energy scales span a wide range from keV down to meV. Therefore, depending on the physics of interest, various effective models can be constructed by integrating out the unrelated degrees of freedom above a cut-off energy~\cite{coleman_intro_2016}. A textbook example is the derivation of the Heisenberg exchange model with only the spin degree of freedom (DOF) as an effective low-energy theory of the Hubbard model that involves both the spin and electron DOF~\cite{fazekas_2003_lecture}. In some special cases, intriguing effective models might exist even if the fundamental DOF stays the same. The dipolar spin-ice state, for example, involves only the spin DOF, but its low-energy dynamics can be explained by a Coulomb gas model of magnetic monopoles~\cite{castelnovo_magnetic_2008, fennell_magnetic_2009, gao_dipolar_2018}.  

Since a hierarchy in the coupling strengths naturally separates modes of different characteristic energies, magnetic systems composed of alternating bonds offer an incomparable opportunity to achieve novel spin correlations that are manifested through effective models~\cite{gao_ordering_2019, kimura_experimental_2014, rau_anisotropic_2016, haku_low_2016,janson_quantum_2014, romhanyi_entangled_2014}. Clusters with a relatively rigid spin alignment could emerge over the stronger bonds and behave as single entities below a cut-off energy. Above this cut-off energy, excitations within the clusters become active, and their propagation is able to account for the dispersion at high energies~\cite{romhanyi_entangled_2014, portnichenko_magnon_2016, tucker_spin_2016}.

\begin{figure}[t!]
\includegraphics[width=0.48\textwidth]{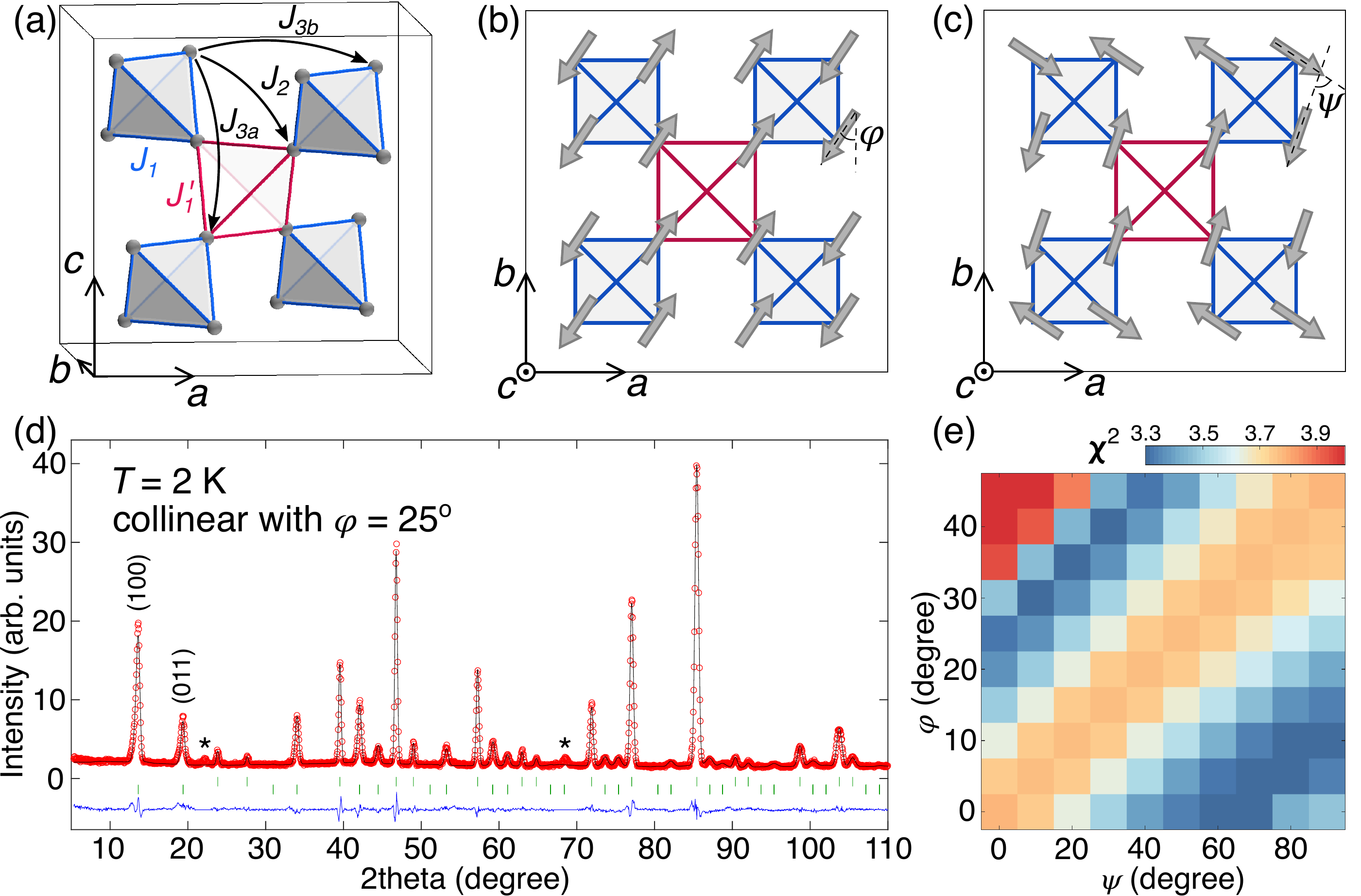}
\caption{(a) The breathing pyrochlore lattice formed by the Cr$^{3+}$ spins (shown as gray spheres) in CuInCr$_4$S$_8$. The complete crystal structure can be found in the Supplemental Materials~\cite{supp}. The bonds of the smaller and larger tetrahedra are shown in blue and red, respectively. Further-neighbor couplings including the second-neighbor coupling $J_2$ and two third-neighbor couplings $J_{3a}$ and $J_{3b}$ are indicated. (b) Collinear magnetic structure viewed along the $c$ axis. $\varphi$  denotes the tilting angle from the $b$ axis. (c) Coplanar magnetic structure that is composed of two AF spin pairs over the smaller tetrahedra. $\psi$ denotes the angle between the two spin pairs. (d) Refinement result of the powder neutron diffraction data measured on HB-2A at $T = 2$~K assuming a collinear structure with $\varphi = 25^{\circ}$. Data were collected with a constant incident neutron wavelength of 2.41 {\AA}. The impurity peaks marked by stars are excluded in the refinements. Data points are shown as red circles. The calculated pattern is shown as the black solid line. The vertical bars indicate the positions of the structural (upper) and magnetic (lower) Bragg peaks for CuInCr$_4$S$_8$. The blue line at the bottom shows the difference of measured and calculated intensities. Similarly good refinement is obtained for nonzero $\psi$, with the refined moment magnitude staying constant at 2.54(2)~$\mu_\textrm{B}$. (e) Map for the goodness-of-fit, $\chi^2$, as a function of $\psi$ and $\varphi$.
\label{fig:pnd}}
\end{figure}

The focus of our study is the spin dynamics in the $A$-site ordered spinel CuInCr$_4$S$_8$, where the magnetic Cr$^{3+}$ ions ($S=3/2$) form a breathing pyrochlore lattice of alternating bond lengths  as shown in Fig.~\ref{fig:pnd}(a)~\cite{pinch_new_1970}. Due to the disparate contributions from the direct exchange interactions that are highly distance sensitive~\cite{gorkom_optical_1973, lee_local_2000}, the coupling strengths over the neighboring tetrahedra denoted as $J_1$ and $J_1'$ are expected to be very different~\cite{okamoto_breathing_2013,ghosh_breathing_2019}. The tetrahedra with the stronger exchange interactions therefore tend to form effective local spin clusters. In the isomorphic compound LiGaCr$_4$S$_8$, clustering over the $J_1'$-bonded tetrahedra have been observed, although the spin-glass-like ground state prevents a quantitative analysis of the spin excitations~\cite{pokharel_negative_2018, pokharel_cluster_2020}. CuInCr$_4$S$_8$ is known to enter an antiferromagnetic (AF) long-range ordered state below $T_N\sim35$~K~\cite{plumier_magnetic_1971, okamoto_magnetic_2018}, which leads to coherent excitations that can be directly compared against effective cluster models.

In a previous neutron diffraction measurement of CuInCr$_4$S$_8$, spins over the $J_1'$-bonded tetrahedra were found to be parallel~\cite{plumier_magnetic_1971}, which has been attributed to the alternating AF $J_1$ and ferromagnetic (FM) $J_1'$ couplings~\cite{ghosh_breathing_2019, benton_ground_2015}. Therefore, as long as the $J_1'$ couplings are sufficiently strong, the low-energy dynamics in CuInCr$_4$S$_8$ can be approximated by correlated tetrahedra with FM spins. However, the Curie-Weiss temperature of $\Theta_{CW} \sim -70$~K suggests dominant AF couplings~\cite{okamoto_magnetic_2018}, which may invalidate a FM tetrahedron model. Further complications could arise from spin-lattice coupling that is often observed in the Cr-based spinels~\cite{lee_local_2000, gao_manifolds_2018}. Recent high-field measurements on CuInCr$_4$S$_8$ revealed a half-magnetization plateau that is reminiscent of a similar feature observed in ZnCr$_2$O$_4$, indicating the existence of biquadratic interactions that could arise from spin-lattice coupling~\cite{okamoto_magnetic_2018, gen_magnetization_2020, penc_half_2004, bergman_models_2006}.

In this paper, we investigate the spin dynamics in CuInCr$_4$S$_8$ using neutron scattering and show the emergence of hierarchical excitations that can be described by an effective model of correlated tetrahedra. In addition to the spin excitations, we observe a strong magnon-phonon coupling at high energies that is further confirmed through the density functional theory (DFT) calculations. The coupling to the lattice DOF provides a possible explanation for the enhanced stability of the spin tetrahedra.

Figure~\ref{fig:pnd} summarizes the refinement results of our powder neutron diffraction data collected on HB-2A at the High Flux Isotope Reactor (HFIR) at ORNL. Details for the sample preparation, basic characterizations, and experimental and theoretical methods can be found in the Supplemental Materials~\cite{supp}. The crystal symmetry of CuInCr$_4$S$_8$ stays cubic  (space group $F\overline{4}3m$) in our investigated temperature range of $T$ = 2~-~300~K, and the Cr-Cr bond distances at $T=2$~K are refined to be 3.37(1) and 3.73(1) Å for $J_1$ and $J_1'$, respectively. Magnetic reflections belonging to the propagation vector of $\bm{q} = (100)$ emerge below $\sim 35$~K. Assuming the magnetic order is described by a single basis vector of the irreducible representation $\Gamma_3$~\cite{supp}, the best refinement is achieved by a collinear structure shown in Fig.~\ref{fig:pnd}(b) with spins perpendicular to $\bm{q}$, \textit{i.e.} $\varphi=0$. Anti-parallel (parallel) spins are observed over the $J_1$-bonded ($J_1'$-bonded) tetrahedra, consistent with the scenario of alternating AF-FM couplings~\cite{plumier_magnetic_1971, benton_ground_2015, ghosh_breathing_2019}. However, as detailed in the Supplemental Materials~\cite{supp}, this solution overestimates the intensity of the (100) reflection, suggesting the necessity to include more than one basis vector.

In order to improve the refinement, we explore different coplanar magnetic structures satisfying two constraints of a $J_1$-$J_1'$ model~\cite{benton_ground_2015}: (1) spins over the $J_1'$-bonded tetrahedra are parallel as required by the FM couplings; (2) spins over the $J_1$-bonded tetrahdera form two AF pairs, leading to a total zero moment as required by the AF couplings. Without loss of generality, we further assume the spins are within the $ab$-plane that contains the $\bm{q}$ direction, thus leaving only three parameters for the refinement: the spin magnitude, the $\psi$ angle between the two AF pairs over the $J_1$-bonded tetrahdera, and the $\varphi$ angle describing an overall rotation with respect to the $c$ axis. Typical refinement results are shown in Fig.~\ref{fig:pnd}(d), where the rotated collinear structure provides a good fit for the diffraction data, and the reduced-$\chi^2$ goodness-of-fit factor is improved from 3.77 at $\varphi = 0$ to 3.31 at $\varphi=25^{\circ}$. As summarized in Fig.~\ref{fig:pnd}(e), equally good refinements are obtained for different $\psi$ values, indicting our diffraction data on a powder sample cannot resolve the AF alignment over the $J_1$-bonded tetrahedra. 

\begin{figure}[t!]
\includegraphics[width=0.4\textwidth]{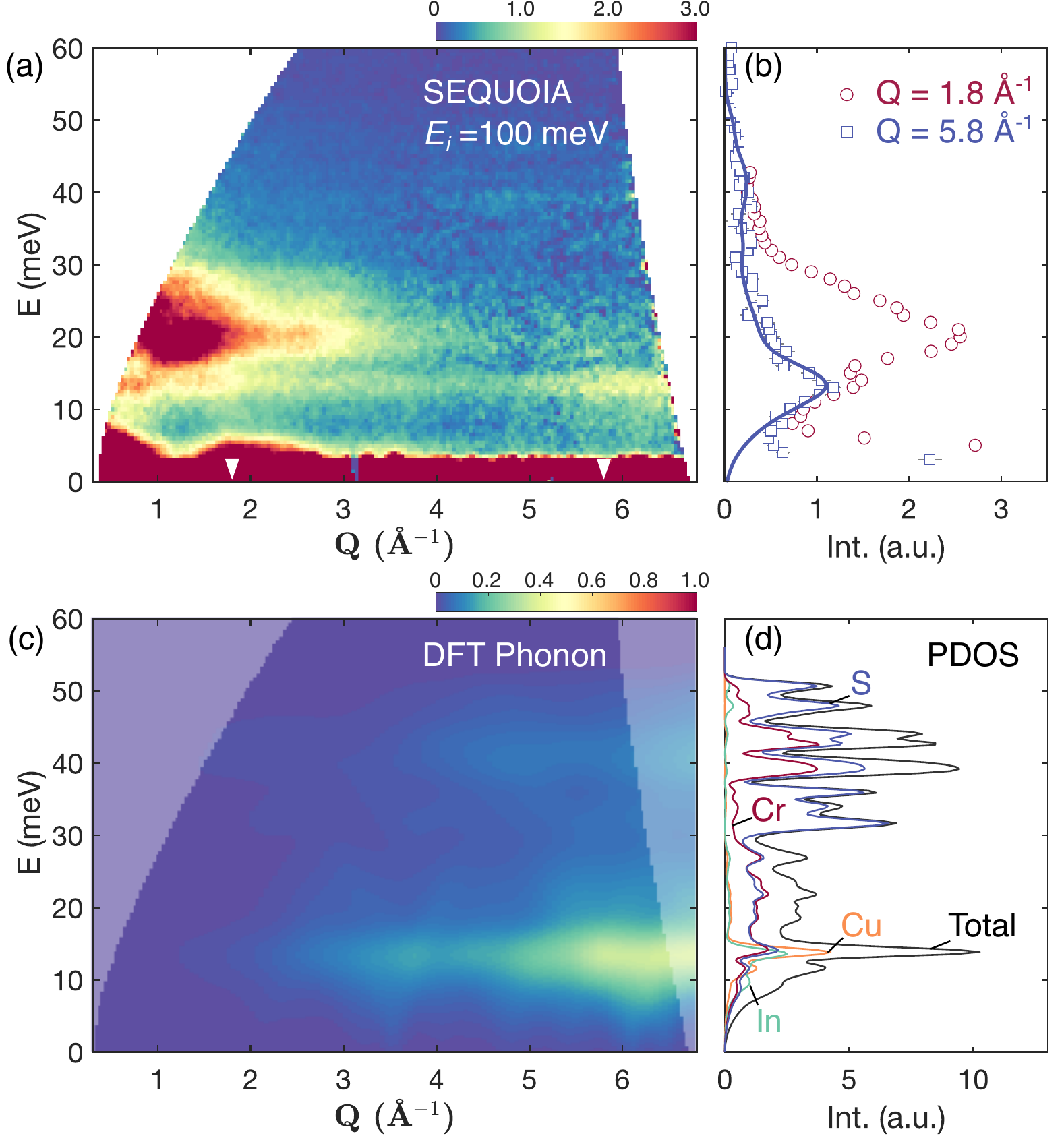}
\caption{(a) Scattering intensity of CuInCr$_4$S$_8$ measured with $E_i = 100$~meV at $T=5$~K. (b) Scattering intensity as a function of $E$ integrated at $Q =1.8$~\AA$^{-1}$ (red circles) and 5.8~\AA$^{-1}$ (blue squares) with an integration width of 0.2~\AA$^{-1}$ as indicated by the white markers in panel (a). Solid line is the phonon scattering intensity at $Q = 5.8$~\AA$^{-1}$ obtained from the DFT calculations after convolution with the instrumental energy resolution. (c) Phonon scattering intensity from the DFT calculations. The calculated data are convoluted with the tabulated instrumental energy resolution. (d) Total and projected density of states (PDOS) of phonons in CuInCr$_4$S$_8$ from the DFT calculations. 
\label{fig:phonon}}
\end{figure}

Although some aspects of the magnetic structure remain undetermined, our inelastic neutron scattering (INS) experiments unambiguously reveal the emergence of correlated spin tetrahedra in CuInCr$_4$S$_8$.  Figure~\ref{fig:phonon}(a) is an overview of the INS spectra measured at $T=5$~K with $E_i=100$~meV. Measurements were performed on SEQUOIA at the Spallation Neutron Source (SNS) at ORNL~\cite{supp}. At low wave-vector transfer, $Q$, hierarchical magnetic excitations are observed in two well-separated energy regimes of [0, 5]~meV and [12, 30]~meV, indicating possible spin-cluster dynamics in CuInCr$_4$S$_8$. In the high-$E$ regime, intensity of the $12$~meV mode is strong on both the low-$Q$ and high-$Q$ sides as confirmed in the constant-$Q$ scans shown in Fig.~\ref{fig:phonon}(b), which suggests a strong coupling between the magnon and phonon excitations~\cite{oh_spontaneous_2016, toth_electromagnon_2016}. The contributions from the phonon excitations are further corroborated in our DFT calculations. The simulated neutron scattering cross section for lattice excitations shown in Fig.~\ref{fig:phonon}(c) reproduces the INS spectra at high $Q$, where the strong intensity at $\sim12$~meV can be ascribed to the high phonon density of states as shown in Fig.~\ref{fig:phonon}(d).

\begin{figure}[t!]
\includegraphics[width=0.48\textwidth]{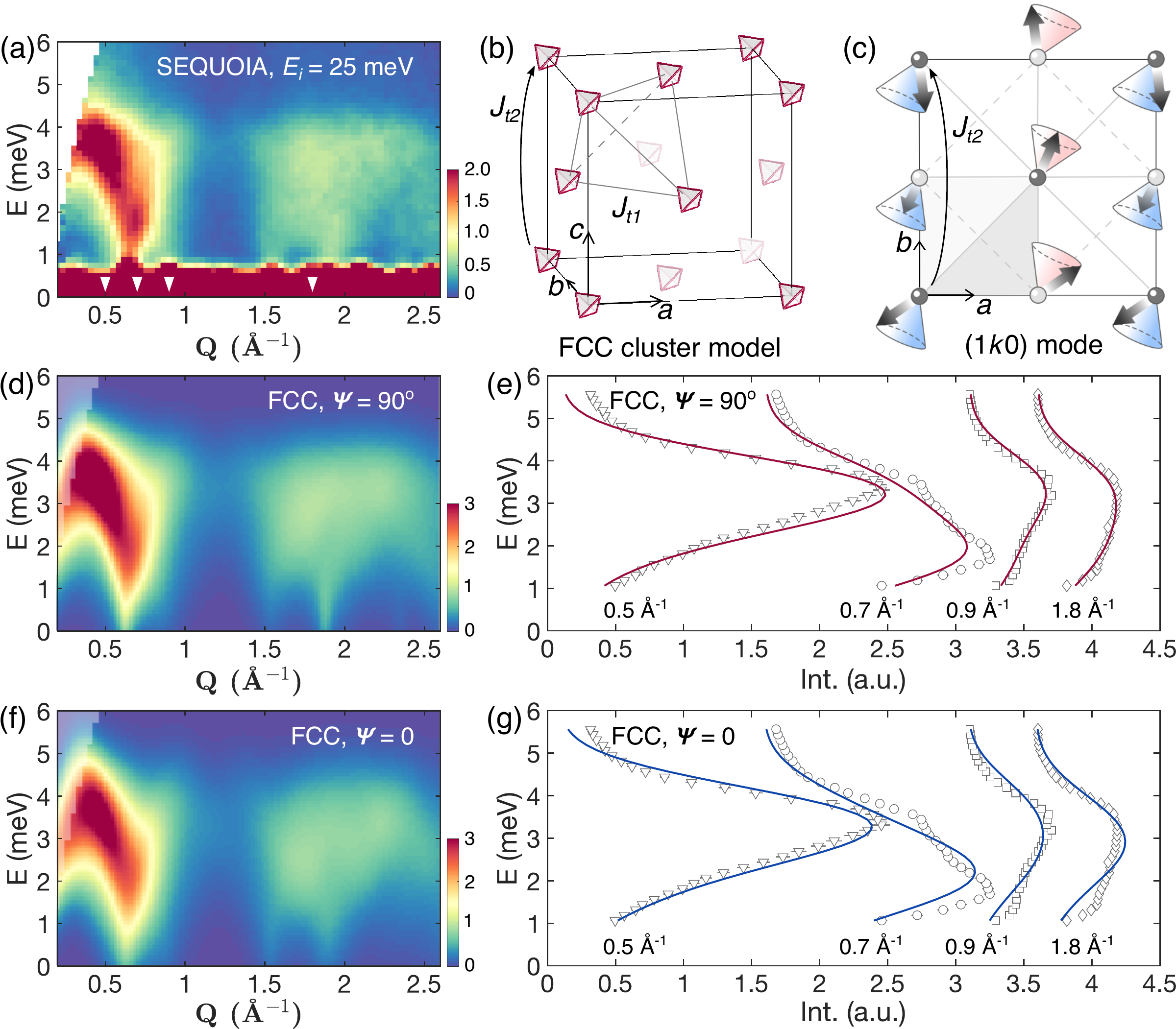}
\caption{(a) Scattering intensity of CuInCr$_4$S$_8$ measured with $E_i = 25$~meV at $T=5$~K. (b) Effective FCC cluster model where the $J_1'$-bonded tetrahedra are treated as rigid FM clusters. The nearest- and second-neighbor couplings among the tetrahedra are indicated as $J_{t1}$ and $J_{t2}$, respectively. (c) Zero-energy mode along the (1$q$0) direction in the FCC cluster model with the inter-tetrahedron couplings limited to the nearest neighbors, \textit{i.e.} $J_{t2}=0$. (d) Intensity calculated by linear spin wave theory using the FCC cluster model with fitted coupling strengths of $J_{t1} = 0.099(9)$~meV and $J_{t2} = -0.022(2)$~meV. The coplanar structure with $\psi = 90^{\circ}$ as shown in Fig.~\ref{fig:pnd}(c) is assumed as the ground state. The calculated data are convoluted with a Gaussian function with a fitted full width at half maximum (FWHM) of 1.6(1)~meV. (e) The corresponding fits for the constant-$Q$ scans that are integrated at $Q=0.5$, 0.7, 0.9, and 1.8~\AA$^{-1}$ with an integration width of 0.1~\AA$^{-1}$ as indicated by the white markers in panel (a). Data at 0.7, 0.9, and 1.8~\AA$^{-1}$ are shifted horizontally by 1.5, 3.0, and 3.5 units for clarity of the figure, respectively. The reduced-$\chi^2$ factor for the fit is 7.4.  (f, g) Similar as panels (d, e) but assuming a collinear ground state shown in Fig.~\ref{fig:pnd}(b). The fitted coupling strengths are $J_{t1} = 0.074(7)$~meV and $J_{t2} = -0.035(3)$~meV. The reduced-$\chi^2$ factor for the fit is 13.2.
\label{fig:cluster}}
\end{figure}

Given the hierarchy of excitations in CuInCr$_4$S$_8$, we can select a cut-off energy between 5 and 12 meV and compare the low-energy spectra shown in Fig.~\ref{fig:cluster}(a) against different effective models. In contrast with the excitations in many regular pyrochlore lattice compounds~\cite{hallas_universal_2016, guratinder_multi_2019}, the low-energy excitations in CuInCr$_4$S$_8$ exhibit a strong modulation along $Q$ that is reminiscent of the cluster-like excitations in MgCr$_2$O$_4$~\cite{tomiyasu_molecular_2008}. For a pair of FM spins at distance $d$, the powder-averaged scattering intensity is modulated by the spherical Bessel function $j_0(Qd) = \frac{\sin(Qd)}{Qd}$~\cite{haraldsen_neutron_2005, houchins_generalization_2015, supp}. Therefore, the minimal intensity observed at $Q\sim1.2$~\AA$^{-1}$ in Fig.~\ref{fig:cluster}(a) implies strong FM correlations at $d\sim3.7$~\AA\ that is close to the $J_1'$ bond length of 3.73(1)~\AA, suggesting a possible effective model of FM tetrahedra over the $J_1'$-bonds.

To quantitatively analyze the low-energy excitations, we extend the previously proposed static cluster model to the dynamical regime~\cite{ghosh_breathing_2019, pokharel_cluster_2020, supp}. The Heisenberg Hamiltonian on the breathing pyrochlore lattice can be written as $\mathcal{H} = \sum_{\langle ij \rangle \in n} J(\bm{r}_{ij})\bm{S}_i\bm{S}_j$ with $n$ listing the coupled spin pairs separated by $\bm{r}_{ij}$. Under modified bond distances $\bm{r}_{ij}'$ = $\bm{r}_{ij}+\delta \bm{r}_{ij}$, the magnon dispersion stays unchanged because the Fourier transform of $J(\bm{r}_{ij})$ and $J(\bm{r}_{ij}')$ in reciprocal space are related by a unitary transformation that does not affect the eigenvalues~\cite{supp}. This freedom in bond distances allows us to treat the original breathing pyrochlore lattice as a face-centered cubic (FCC) lattice of point-like tetrahdra as shown in Fig.~\ref{fig:cluster}(b). Under the rigid cluster approximation, the tetrahedra become composite spins of $S = 6$. The intra-tetrahedron couplings $J_1'$ are thus integrated out, and the remaining Cr-Cr couplings up to the third-neighbors shown in Fig.\ref{fig:pnd}(a) are merged into an effective coupling $J_{t1}=(J_1+4J_2+2J_{3a}+2J_{3b})/16$ over the nearest-neighbor tetrahedra as presented in Fig.~\ref{fig:cluster}(b).

\begin{figure}[t!]
\includegraphics[width=0.48\textwidth]{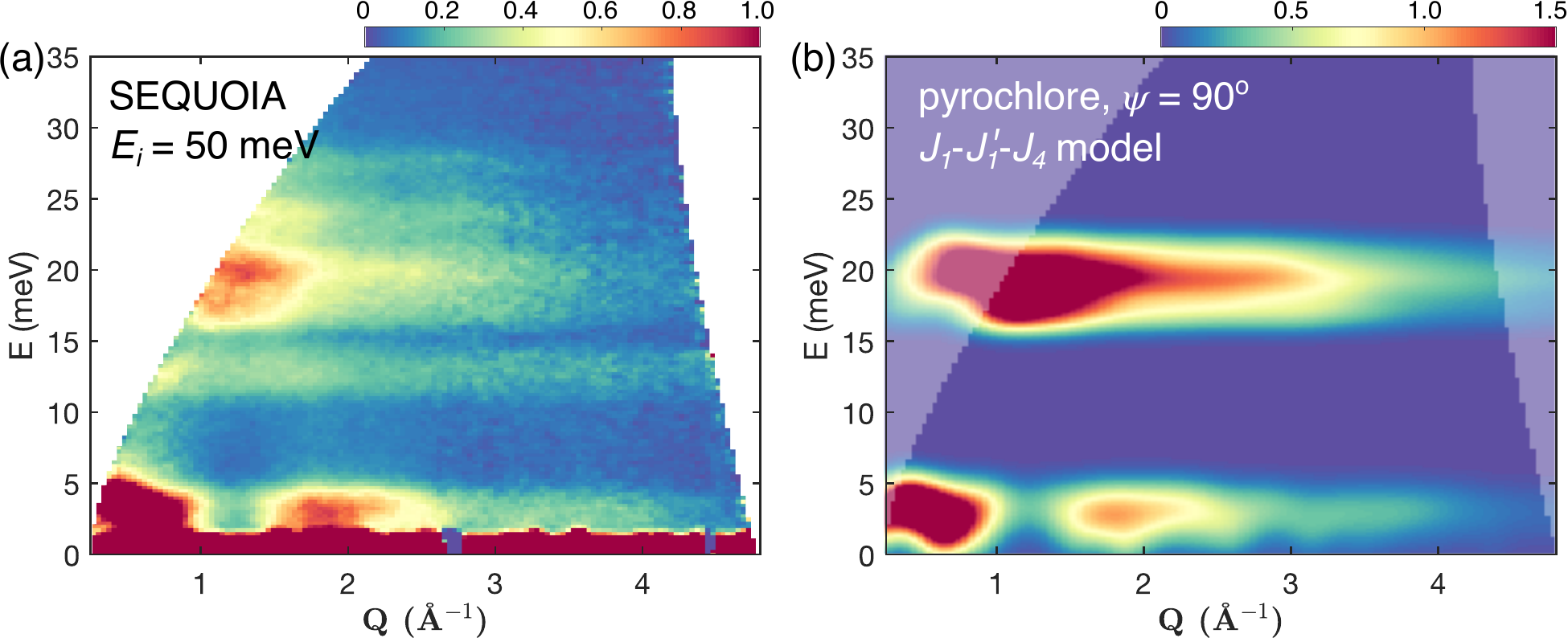}
\caption{(a) Scattering intensity of CuInCr$_4$S$_8$ measured with $E_i = 50$~meV at $T=5$~K. (b)  Intensity map calculated by the linear spin wave theory using the minimal $J_1$-$J_1'$-$J_4$ model on the breathing pyrochlore lattice. The strength of $J_1'$ is fixed at $-2.6$~meV to reproduce the central positions of the high-$E$ modes. For the coplanar ground state with $\psi=90^{\circ}$, the coupling strengths obtained by fitting the low-energy spectra are $J_1 = 1.8(1)$~meV, $J_4 = -0.11(1)$~meV, close to the expected values of $J_1 = 16J_{1t}$ and $J_4 = 4J_{t2}$ of the FCC cluster model~\cite{supp}. The calculated data are convoluted by a Gaussian function with a FWHM that is two times broader than the tabulated instrumental energy resolution. Similar results can be obtained for the ground states with different $\psi$.  
\label{fig:highE}}
\end{figure}

The magnon dispersion of the FCC tetrahedron model with couplings up to the second neighbors ($J_{t2}$) is then calculated using linear spin wave theory~\cite{toth_linear_2015}. The scattering intensities at wave-vector transfer $\bm{k}$ are scaled by the structure factor of a $J_1'$-bonded tetrahedron $A_{\rm{tetra}}(\bm{k}) = \sum_{i=1}^4 e^{-i\bm{kr}_i}$ to recover the actual spin distribution on the original pyrochlore lattice~\cite{supp}. For the fits of the INS spectra, constant-$Q$ scans at $Q=0.5$, 0.7, 0.9, and 1.8~\AA$^{-1}$ are generated as the inputs, shown in Fig.~\ref{fig:cluster}(e) and (g). Starting from the coplanar ground state with $\psi = 90^{\circ}$ as shown in Fig.~\ref{fig:pnd}(c), the INS spectra are reproduced with the coupling strengths $J_{t1} = 0.099(9)$~meV and $J_{t2} = -0.022(2)$~meV as summarized in Figs.~\ref{fig:cluster}(d) and (e). The calculated spectra only depends weakly on the $\psi$ angle in the ground state. As exemplified in Figs.~\ref{fig:cluster}(f) and (g), the main features of the INS spectra are reproduced even with a collinear ground state ($\psi=0$) with $J_{t1} = 0.074(7)$~meV and $J_{t2} = -0.035(3)$~meV. Thus the FCC tetrahedron model provides a good approximation for the low-energy dynamics in CuInCr$_4$S$_8$ regardless of the ambiguity in the ground state.

The effective $J_{t2}$ couplings can be ascribed to the Cr-Cr couplings beyond the third neighbors on the original pyrochlore lattice. Its FM character is consistent with the propagation vector $\bm{q}=(100)$ as it favors parallel spin alignment across the neighboring unit cells. Without $J_{t2}$, zero-energy modes could emerge due to the geometric frustration of the FCC lattice~\cite{balla_degenerate_2020}, which results in a worse fit of the INS spectra~\cite{supp}. Assuming a collinear ground state as shown in Fig.~\ref{fig:pnd}(b), Fig.~\ref{fig:cluster}(c) plots a snapshot for one of the zero-energy modes at $\bm{k}=(1\frac{1}{4}0)$. The FM $J_{t2}$ coupling raises this mode to a nonzero energy and partially relieves the ground state degeneracy, thus fostering the magnetic long-range order in CuInCr$_4$S$_8$.

The spin dynamics at higher energies involve excitations within the FM tetrahedra. For free tetrahedra, neutron scattering detects the transition from the ground state to the first excited state at $E=6J'$, and the powder-averaged scattering cross section is proportional to $1-j_0(Qd)$~\cite{haraldsen_neutron_2005, supp}. Therefore, at high energies, stronger intensity is expected at $Q\sim1.2$~\AA$^{-1}$, which is indeed observed in our INS spectra at $\sim19$~meV as shown in Fig.~\ref{fig:highE}(a). The split branches at $\sim 17$ and 20~meV can be ascribed to the bottom and top of the unresolved dispersion, respectively. As shown in Fig.~\ref{fig:highE}(b), with a strong $J_1'$ coupling of $-2.6$~meV, the split branches can be reproduced using the minimal $J_1$-$J_1'$-$J_4$ model on the original pyrochlore lattice. The obtained $J_1'$ value is consistent with the tetrahedron spectra and explains the hierarchical excitations in CuInCr$_4$S$_8$.

Despite the success of the effective low-energy model, there are still several outstanding questions concerning the physical behavior of CuInCr$_4$S$_8$. In particular the INS spectra shown in Fig.~\ref{fig:highE} exhibit additional
high-energy modes compared to the anticipation of the minimal model. Moreover, the negative Curie-Weiss temperature $\Theta_{CW}$ seems at odds with the dominating FM $J_1'$ interactions. Both phenomena may be related to the coupling to the lattice DOF as discussed in conjunction with Fig.~\ref{fig:phonon}. Along with interactions neglected in our minimal model, the coupling to the lattice DOF could produce extra modes that is beyond a spin-only description. Additionally, the recent high-field magnetization measurements~\cite{okamoto_magnetic_2018, gen_magnetization_2020} demonstrate that the spin-lattice coupling in CuInCr$_4$S$_8$ induces effective biquadratic interactions, which provide a possible explanation of the negative $\Theta_{CW}$ as biquadradratic interactions are known to enhance the Heisenberg (billinear) interactions between the FM spins without affecting $\Theta_{CW}$~\cite{wysocki_consistent_2011, wildes_evidence_2020}. Therefore, the strong FM $J_1'$ in CuInCr$_4$S$_8$ may receive a considerable enhancement through the coupling to the lattice DOF. Further studies on a single crystal sample will help elucidate the role of the spin-lattice coupling in CuInCr$_4$S$_8$.

Our analysis of the hierarchical excitations in CuInCr$_4$S$_8$ exemplifies a general effective-cluster approach to understand the spin dynamics on the breathing-type lattices, where a series of exotic states like Weyl magnons~\cite{li_weyl_2016}, magnetic skyrmions~\cite{hirschberger_skyrmion_2019, matsumura_helical_2019}, and spiral spin liquids~\cite{ghosh_breathing_2019, bergman_order_2007, gao_spiral_2017, gao_fractional_2020} have been realized or proposed. Even for compounds without hierarchical excitations, an effective cluster model could still provide a good approximation for the quasi-elastic dynamics as long as the intra-cluster excitations are separated from the ground state by a nonzero gap. Conversely, by decorating a regular lattice with clusters, models with different high-energy dynamics could be designed without altering the low-energy responses, which might be utilized in tailoring the properties of functional materials~\cite{tokura_emergent_2017}.

In conclusion, our neutron scattering study of CuInCr$_4$S$_8$ reveals hierarchical excitations that can be described by a FCC lattice of correlated tetrahedra. The magnon-phonon coupling observed in our INS spectra does not affect the spin dynamics at low energies where the spins in the tetrahedra are approximately parallel. Instead, the coupling to the lattice DOF may enhance the stability of the spin tetrahedra. The effective cluster model demonstrated in our work can be generalized to different breathing-type lattice compounds and help understand their spin dynamics.

\begin{acknowledgments}
We acknowledge helpful discussions with K.D. Belashchenko, Y.Q. Cheng, M. McGuire, and H.B. Cao. This work was supported by the U.S. Department of Energy, Office of Science, Basic Energy Sciences, Materials Sciences and Engineering Division. This research used resources at the Spallation Neutron Source (SNS) and the High Flux Isotope Reactor (HFIR), both are DOE Office of Science User Facilities operated by the Oak Ridge National Laboratory (ORNL). H.S.A and G.P. were supported by the Gordon and Betty Moore Foundation’s EPiQS Initiative through Grant GBMF4416. 
\end{acknowledgments}

\clearpage
\newpage

\renewcommand{\thefigure}{S\arabic{figure}}
\renewcommand{\thetable}{S\arabic{table}}
\renewcommand{\theequation}{S\arabic{equation}}

\makeatletter
\renewcommand*{\citenumfont}[1]{S#1}
\renewcommand*{\bibnumfmt}[1]{[S#1]}
\def\clearfmfn{\let\@FMN@list\@empty}  
\makeatother
\clearfmfn

\setcounter{figure}{0} 
\setcounter{table}{0}
\setcounter{equation}{0} 

\onecolumngrid
\begin{center} {\bf \large Hierarchical excitations from correlated spin tetrahedra on the breathing pyrochlore lattice \\
 Supplementary Information} \end{center}

\vspace{0.5cm}

\section{Experimental and Theoretical Methods}
Polycrystalline CuInCr$_4$S$_8$ was synthesized by a direct reaction of the high-purity elements in evacuated silica ampoules. The ampoules were heated slowly (25-30 $^{\circ}$C/h) to an ultimate temperature of 900$^{\circ}$C for 48-72~h, with an intermediate dwell of 24-48~h at 600$^{\circ}$C. Following the initial reaction, the material was ground in air, pressed into a pellet, sealed in an evacuated ampoule and again heated to 900$^{\circ}$C for up to 4 days.

Neutron diffraction experiments were performed on POWGEN at the Spallation Neutron Source (SNS) and HB-2A at the High Flux Isotope Reactor (HFIR) at ORNL. A powder sample with a total mass of $\sim 7$~g was placed in a vanadium sample can. On POWGEN, data were acquired between 10 and 300 K using the 0.8 and 2.67 {\AA} instrumental configurations. On HB-2A, data were acquired at $T$ between 2 and 160 K with a constant incident neutron wavelength of 2.41 {\AA}. Rietveld refinements of the neutron diffraction data were performed using the FULLPROF program~\cite{rodriguez_recent_1993}. Refinement results of the POWGEN data are similar to those of the HB-2A data and are thus not shown.

Inelastic neutron scattering (INS) experiments were performed on SEQUOIA at the SNS. The same powder sample was placed in an aluminum can. A closed cycle refrigerator (CCR) was employed to reach temperatures $T$ down to 5 K. The incident neutron energy was selected at $E_i$ = 10, 25, 50, and 100 meV, with a corresponding Fermi chopper frequency of 180, 240, 360, and 540 Hz, respectively in the high-resolution setup. Data from an empty sample can was subtracted as the background. Data reductions were performed using the MANTID software~\cite{arnold_mantid_2014}. Linear spin wave calculations were performed using the SpinW program~\cite{toth_linear_2015s}.

The phonon density of states (DOS) of CuInCr$_4$S$_8$ was calculated based on the frozen phonon method using the PHONOPY code~\cite{togo_first_2015} by assuming a FM ordered state. Forces were calculated based on DFT implemented in the VASP code~\cite{kresse_efficiency_1996}. The DFT+U method~\cite{dudarev_electron_1998} was used with $U – J = 6$~eV~\cite{isseroff_importance_2012} and 3.7~eV~\cite{wang_oxidation_2006} applied on the Cu-3d and Cr-3d orbitals, respectively. The interaction between ions and electrons was described by the projector augmented plane-wave method~\cite{kresse_from_1999}. A kinetic energy cutoff of 400 eV was used for the plane-wave basis. The calculated data are convoluted with the tabulated instrumental energy resolution using the OCLIMAX program~\cite{cheng_simulation_2019}.

\begin{figure}[b!]
\includegraphics[width=0.95\textwidth]{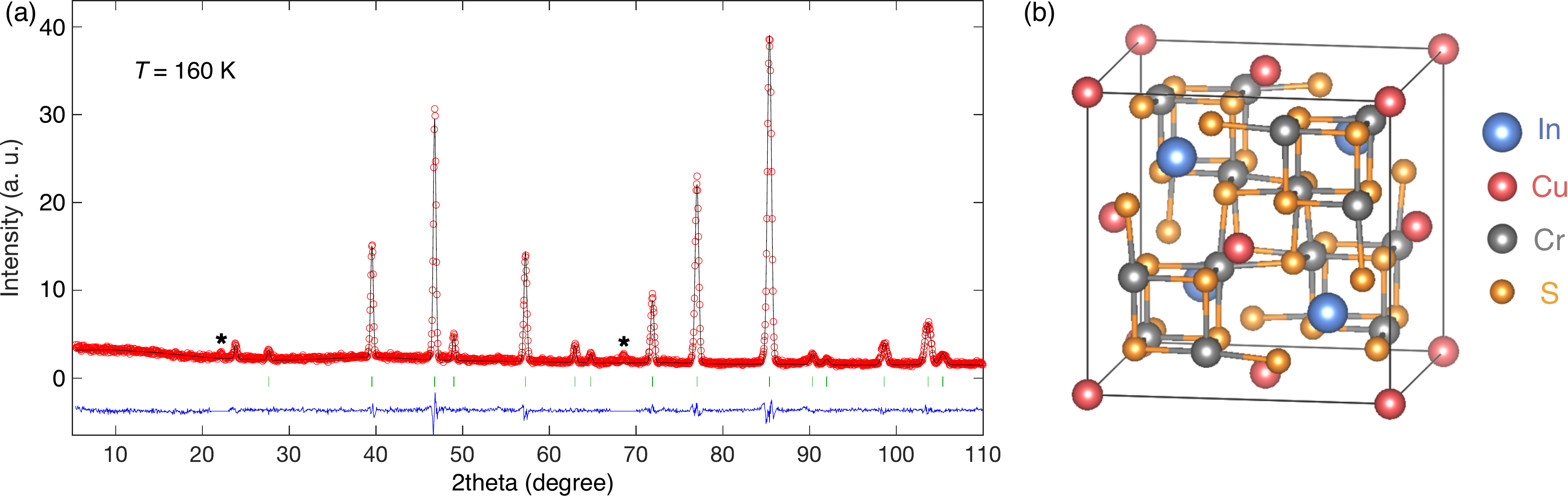}
\caption{(a) Refinement result of the powder neutron diffraction data measured on HB-2A at $T = 160$~K.  The vertical bars indicate the positions of the structural Bragg peaks for CuInCr$_4$S$_8$. Data points and calculated pattern are presented similarly as that in Fig.~1(d) of the main text. (b) Crystal structure of CuInCr$_4$S$_8$ with the Cr-S bonds shown explicitly.
\label{fig:nuclear}}
\end{figure}

\section{Sample quality verified by powder neutron diffraction}
The quality of our CuInCr$_4$S$_8$ sample was checked using powder neutron diffraction. Figure~\ref{fig:nuclear} plots the refinement result of the diffraction data collected on HB-2A at $T=160$~K. With the cubic space group $F\overline{4}3m$ (\#216), the lattice constant was refined to be $a = 10.0414(2)$~\AA. The Cr ions occupy the $16e$ $(x,x,x)$ Wyckoff sites with $x = 0.3700(5)$. The Cu and In ions occupy the $4a$ $(0,0,0)$ and $4d$ $(\frac{3}{4},\frac{3}{4},\frac{3}{4})$ sites, respectively. The S ions occupy two $16e$ $(x,x,x)$ sites with $x = 0.1330(5)$ and 0.6109(5). The goodness-of-fit factor is $\chi^2 = 2.30$. Releasing the site occupancy does not improve $\chi^2$, confirming the good quality of our sample.

\section{Magnetic structures from representation analysis}
\begin{figure}[t]
\includegraphics[width=0.9\textwidth]{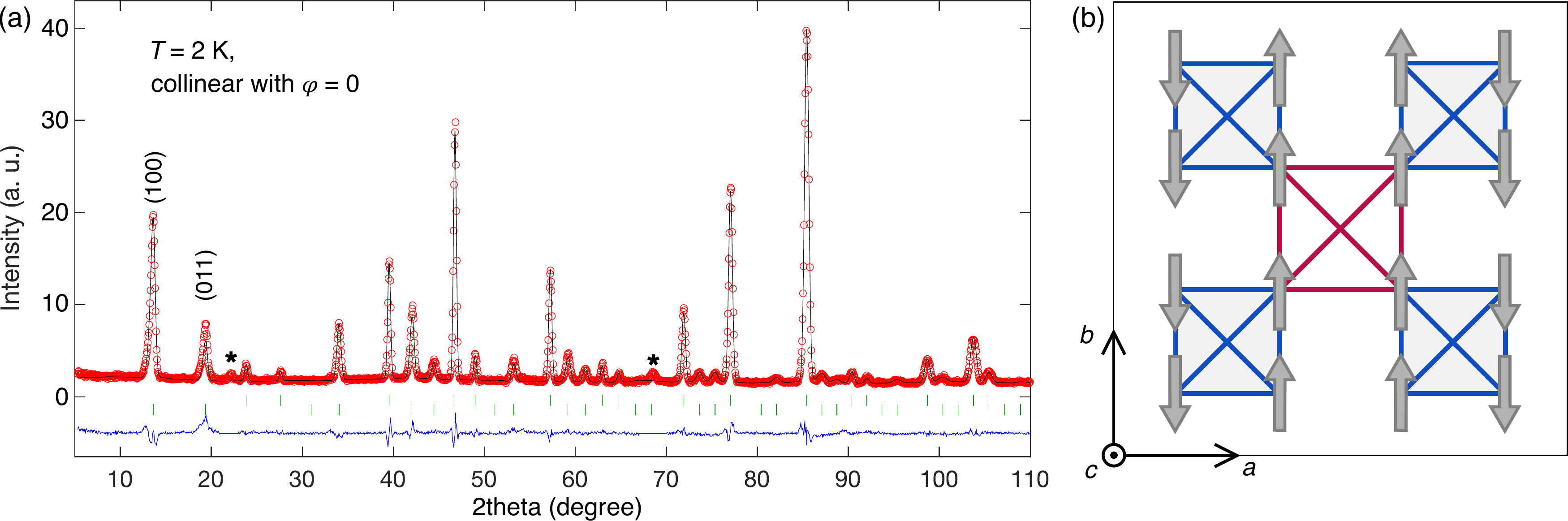}
\caption{(a) Refinement result of the powder neutron diffraction data measured on HB-2A at $T = 2$~K using the single basis vector collinear structure ($\psi = 0$). Data points and calculated pattern are presented similarly as that in Fig.~1(d) of the main text. (b) The collinear magnetic structure with $\psi = 0$.
\label{fig:bv}}
\end{figure}

Representation analysis was performed using the BasIreps program in FullProf~\cite{rodriguez_recent_1993}. For $\bm{q}=(100)$, there are 5 irreducible representations with $\sum_{\nu=1}^5 n_\nu \Gamma_\nu^\mu = \Gamma_1^1 + 2\Gamma_2^1 + 3\Gamma_3^2 + \Gamma_4^1 + 2\Gamma_5^1$, where $n_\nu$ is the number of times that the irreducible representation $\Gamma_\nu$ of order $\mu$ appears in the magnetic representation. Among them, $\Gamma_3$ provides the best single basis vector fit of the our diffraction data as shown in Fig.~\ref{fig:bv}(a), and the corresponding structure is presented in Fig.~\ref{fig:bv}(b). As mentioned in the main text, this solution overestimates the intensity ratio between the (100) and (011) reflections, resulting in a higher $\chi^2 = 3.77$ compared to the rotated magnetic structure shown in Fig.~1 of the main text. The reduced intensity of the (100) reflection for the rotated structure can be understood by the fact that neutron scattering only detects the magnetic structure factor components that are perpendicular to $\bm{q}=(100)$.

\section{Neutron scattering cross section of the tetrahedron excitations}
In the main text, the $Q$-dependence of the INS intensity is exploited to distinguish the magnon modes with and without the intra-tetrahedron excitations. Here we present a brief derivation for the characteristic $Q$-dependence of the tetrahedron excitations.

According to the work of Haraldsen \textit{et al.}~\cite{haraldsen_neutron_2005s}, the neutron scattering cross section for the tetrahedron excitations is proportional to the `exclusive structure factor' expressed as
\begin{equation}
    S_{ba}(\bm{k}) = \sum_{\lambda_f}\langle \Psi_i|V_b^{\dagger}|\Psi_f(\lambda_f)\rangle
    \langle \Psi_f(\lambda_f)|V_a|\Psi_i\rangle \textrm{,}
\end{equation}
where $|\Psi_i\rangle$ and $|\Psi_f(\lambda_f)\rangle$ are the initial and final states of the transition, respectively, the subscripts $a$ and $b$ indicate the components along the Cartesian axes, and the vector $\bm{V}$ is a sum of the spin operators:
\begin{equation}
    \bm{V} = \sum_{i=1}^4 \bm{S}({\bm{r}_i})e^{i\bm{k}\bm{r}_i}\textrm{.}
\end{equation}

\begin{figure}[b!]
\includegraphics[width=0.70\textwidth]{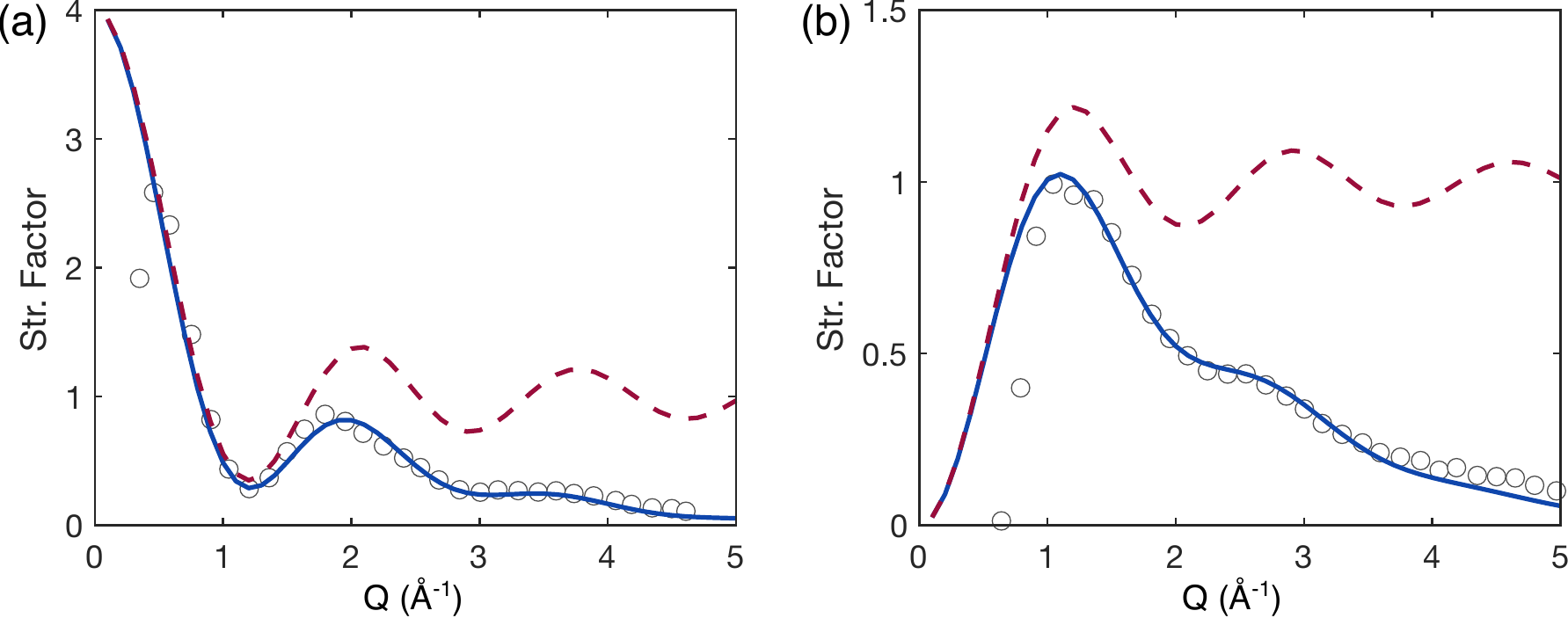}
\caption{The $Q$-dependence of the exclusive structure factor for a FM tetrahedron with $d = 3.73$~\AA~(red dashed lines) with $S(Q) = 1 + 3j_0(Qd)$ in panel (a) for the static correlation and $S(Q) = 1-j_0(Qd)$ in panel (b) for the first excitation. The blue solid lines are the structural factor multiplied by the squared Cr$^{3+}$ magnetic form factor. Gray circles are intensities in arb. units integrated in the energy range of [2,~6] meV for the $E_i=50$~meV spectra (a) and [15,~25] meV for the $E_i=100$~meV spectra. Error bars representing the standard deviations are smaller than the marker size. The deviation of the integrated intensities from the free tetrahedron predictions at low $Q$ is due to kinematic constraints of the neutron resulting in a partial region of energy transfer and wave-vector transfer sampled for fixed incident energy neutrons at small wave vectors.
\label{fig:tetramer}}
\end{figure}

Previous works have shown that the INS structure factor of the Cr$^{3+}$ $S=3/2$ cluster can be reduced to the $S=1/2$ analog~\cite{houchins_generalization_2015s, gao_manifolds_2018s}. Without loss of generality, the ground state of the FM $S=1/2$ tetrahedron can be written as $|\textrm{GS}\rangle = |\uparrow\uparrow\uparrow\uparrow \rangle$ with the $i$-th  $|\uparrow\rangle$ denoting the polarized $|S_z=1/2\rangle$ state at site $\bm{r}_i$. For the static structure factor, the only nonzero component is $S_{zz}$:

\begin{align}
    S_{zz}(\bm{k}) &=\langle GS|V_z^\dagger|GS\rangle \langle GS|V_z|GS\rangle \nonumber \\
    &= \frac{1}{4}(\sum_{i=1}^4 e^{-i\bm{k}\bm{r}_i})(\sum_{i=1}^4 e^{i\bm{k}\bm{r}_i}) \nonumber \\
    &=1 + \frac{1}{2}\sum_{<ij>\in n}\cos{(\bm{k}\bm{d_{ij}})}\textrm{,}
\end{align}
where $n = 1\sim6$ lists the spin pairs in the tetrahedron separated by $\bm{d}_{ij}$ of equal lengths as in the main text. For a powder sample, the $Q$-dependence of the structure factor can be obtained through integration
\begin{equation}
    S_{zz}(Q) = \int \frac{d\Omega_{\bm{\hat{k}}}}{4\pi}S_{zz}(\bm{k}) = 1 + 3j_0(Qd)\textrm{.}
\end{equation}
In this expression, $j_0(x)$ is the spherical Bessel function $j_0(x)=\frac{\sin x}{x}$ and $d$ is the distance between the spin pairs. For the $J_1'$-bonded tetrahedra in CuInCr$_4$S$_8$ with $d = 3.73$~\AA, the $Q$-dependence is plotted in Fig.~\ref{fig:tetramer}(a). The minimal $S_{zz}$ value at $Q\sim1.2$~\AA$^{-1}$ accounts for the reduced INS intensity at low energies as observed in Fig.~3 of the main text.

For the first tetrahedron excitation, the final states are:
\begin{align}
    |\Psi_f(1)\rangle &= \frac{1}{2}(|\downarrow \uparrow \uparrow \uparrow\rangle + |\uparrow \downarrow \uparrow \uparrow\rangle - |\uparrow \uparrow \downarrow \uparrow\rangle - |\uparrow \uparrow \uparrow \downarrow\rangle) \nonumber \\
    |\Psi_f(2)\rangle &= \frac{1}{2}(|\downarrow \uparrow \uparrow \uparrow\rangle - |\uparrow \downarrow \uparrow \uparrow\rangle + |\uparrow \uparrow \downarrow \uparrow\rangle - |\uparrow \uparrow \uparrow \downarrow\rangle)  \nonumber \\
    |\Psi_f(3)\rangle &= \frac{1}{2}(|\downarrow \uparrow \uparrow \uparrow\rangle - |\uparrow \downarrow \uparrow \uparrow\rangle - |\uparrow \uparrow \downarrow \uparrow\rangle + |\uparrow \uparrow \uparrow \downarrow\rangle)\textrm{.}
\end{align}
The corresponding exclusive structure factor $S_{zz}$ can be calculated as:
\begin{equation}
    S_{zz}(\bm{k}) = \sum_{\lambda_f = 1}^3 \langle GS|V_z^\dagger|\Psi_f(\lambda_f)\rangle
    \langle \Psi_f(\lambda_f)|V_z|GS\rangle\textrm{,}
\end{equation}
with
\begin{align}
    \langle\Psi_f(1)|V_z|GS\rangle &= \frac{1}{4}(-e^{i\bm{kr}_1} - e^{i\bm{kr}_2} + e^{i\bm{kr}_3} + e^{i\bm{kr}_4}) \nonumber \\
    \langle\Psi_f(2)|V_z|GS\rangle &= \frac{1}{4}(-e^{i\bm{kr}_1} + e^{i\bm{kr}_2} - e^{i\bm{kr}_3} + e^{i\bm{kr}_4}) \nonumber \\
    \langle\Psi_f(3)|V_z|GS\rangle &= \frac{1}{4}(-e^{i\bm{kr}_1} + e^{i\bm{kr}_2} + e^{i\bm{kr}_3} - e^{i\bm{kr}_4})\textrm{.}
\end{align}
It is immediately clear that the $\bm{k}$-dependence of the first excitation of a FM tetrahedron follows the static structure factor of an AF tetrahedron with two spins up and two spins down. For example, for $\lambda_f = 1$, we have
\begin{align}
    S_{zz}^{\lambda_f = 1}(\bm{k}) &= \langle GS|V_z^\dagger|\Psi_f(1)\rangle \langle\Psi_f(1)|V_z|GS\rangle \nonumber \\
     &= \frac{1}{16}[4 + 2\cos(\bm{kd}_{12}) + 2\cos(\bm{kd}_{34}) - 2\cos(\bm{kd}_{13}) - 2\cos(\bm{kd}_{14})- 2\cos(\bm{kd}_{23})- 2\cos(\bm{kd}_{24})]\textrm{,}
\end{align}
which leads to a $Q$-dependence of
\begin{equation}
    S_{zz}^{\lambda_f = 1}(Q) = \frac{1}{4}[1 - j_0(Qd)]\textrm{.}
\end{equation}
Similar results can be derived for $\lambda_f = 2$ and 3, and also for the other nonzero exclusive structural factor components $S_{xx}$ and $S_{yy}$. Therefore, the cross section of the first excitation will exhibit a reversed $Q$-dependence compared to that of the ground state. As shown in Fig.~\ref{fig:tetramer}(b), after multiplying the magnetic form factor of the Cr$^{3+}$ spins, the maximal INS intensity occurs at $Q\sim1.2$~\AA$^{-1}$ as observed in our INS spectra at $E\sim 19$~meV.

\section{Constant magnon dispersion under atomic displacements}
\begin{figure}[h]
\includegraphics[width=0.5\textwidth]{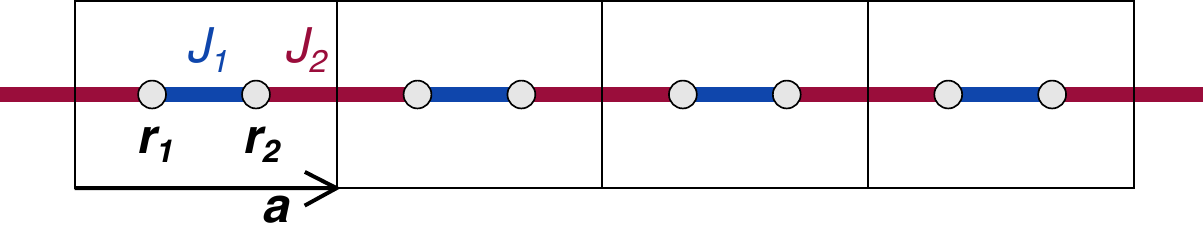}
\caption{$J_1$-$J_2$ spin chain model with two spins at $\bm{r}_1$ and $\bm{r}_2$ in one unit cell. $\bm{a}$ defines the vector along the chain direction with a length equal to the lattice constant.
\label{fig:eigen}}
\end{figure}

Our effective FCC tetrahedron model relies on the fact that magnon dispersion stays constant against atomic displacements. This law can be illustrated using the $J_1$-$J_2$ spin chain model shown in Fig.~\ref{fig:eigen}. With two spins in one unit cell, the Fourier transform of the couplings $J(\bm{r})$ can be written as a $2\times2$ matrix
\begin{equation}
    J(\bm{k}) = 
    \begin{bmatrix}
    0 & J_1e^{i\bm{k}(\bm{r}_2 - \bm{r}_1)} + J_2e^{i\bm{k}(\bm{r}_2 - \bm{a} - \bm{r}_1)} \\
    J_1e^{-i\bm{k}(\bm{r}_2 - \bm{r}_1)} + J_2e^{-i\bm{k}(\bm{r}_2 - \bm{a} - \bm{r}_1)} & 0
    \end{bmatrix}\textrm{.}
\end{equation}
Under an atomic displacement $\bm{r}_1'$ = $\bm{r}_1+\bm{\delta}$, the coupling matrix becomes
\begin{align}
    J'(\bm{k}) &= 
    \begin{bmatrix}
    0 & J_1e^{i\bm{k}(\bm{r}_2 - \bm{r}_1 - \bm{\delta})} + J_2e^{i\bm{k}(\bm{r}_2 - \bm{a} - \bm{r}_1 - \bm{\delta})} \\
    J_1e^{-i\bm{k}(\bm{r}_2 - \bm{r}_1 - \bm{\delta})} + J_2e^{-i\bm{k}(\bm{r}_2 - \bm{a} - \bm{r}_1 - \bm{\delta})} & 0 
    \end{bmatrix} \\
    &= \begin{bmatrix}
    0 & (J_1e^{i\bm{k}(\bm{r}_2 - \bm{r}_1)} + J_2e^{i\bm{k}(\bm{r}_2 - \bm{a} - \bm{r}_1)})e^{-i\bm{k}\bm{\delta}} \nonumber \\
    J_1(e^{-i\bm{k}(\bm{r}_2 - \bm{r}_1)} + J_2e^{-i\bm{k}(\bm{r}_2 - \bm{a} - \bm{r}_1)})e^{i\bm{k}\bm{\delta}} & 0 \\
    \end{bmatrix}\textrm{,}
\end{align}
which is related to $J(\bm{k})$ through a unitary transformation $J'(\bm{k})=U J(\bm{k}) U^{-1}$ with
\begin{equation}
    U = 
    \begin{bmatrix}
    e^{-i\bm{k\delta}} & 0 \\
    0 & 1
    \end{bmatrix}\textrm{.}
\end{equation}

Therefore, the eigenvalues of $J'(\bm{k})$ and $J(\bm{k})$ should be the same as a unitary transformtaion does not affect the eigenvalues. For an ordered state on a more complicated lattice, the magnon dispersion calculated through the Holstein-Primakoff scheme~\cite{toth_linear_2015s} follows similar arguments and also stays constant against atomic displacements. Similar to the conventional magnetic form factor that describes the spin distribution around the magnetic ions, the scattering cross section of the cluster model should be multiplied by the structure factor of the cluster to account for the original spin distribution.

\section{Importance of $J_{t2}$ in describing the INS spectra}
\begin{figure}[h!]
\includegraphics[width=0.95\textwidth]{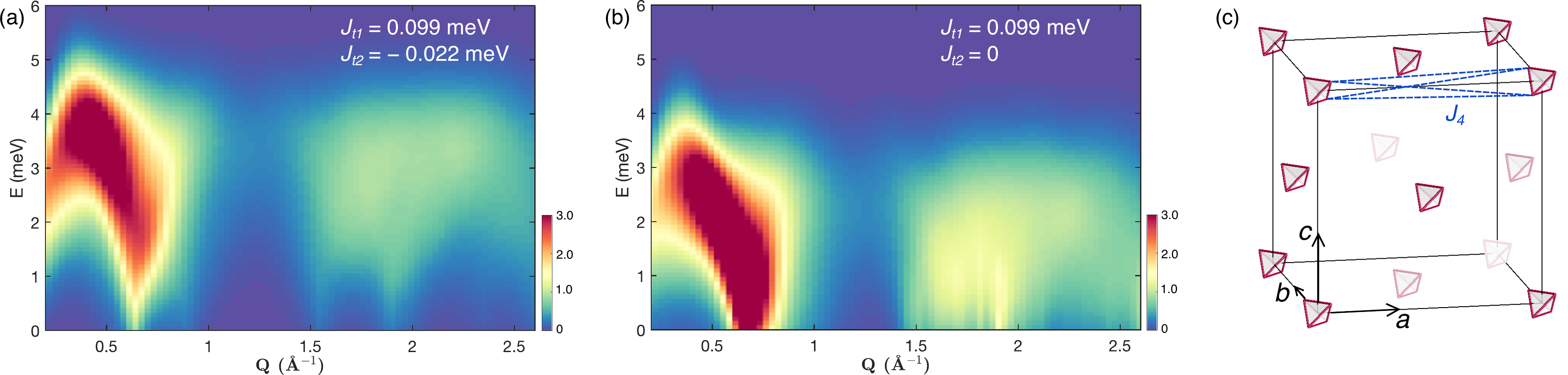}
\caption{INS spectra calculated using the effective FCC model with coupling strengths $J_{t1}=0.099$~meV, $J_{t2}=-0.022$~meV in panel (a) and $J_{t1}=0.099$~meV, $J_{t2}=0$ in panel (b). (c) The 4 exchange paths of $J_4$ (blue dashed lines) that contribute to the effective $J_{t2}$ interactions.
\label{fig:Jt2}}
\end{figure}

With only $J_{t1}$, the zero-energy modes result in strong intensity close to the elastic line as shown in Fig.~\ref{fig:Jt2}(b) compared to Fig~\ref{fig:Jt2}(a). As presented in the main text, FM $J_{t2}$ couplings relieve the degeneracy and improve the fit. This effective $J_{t2}$ can be attributed to the couplings beyond the third-neighbors on the original pyrochlore lattice. For the minimal $J_1$-$J_1'$-$J_4$ model discussed in the main text, we have $J_{t2}=J_4/4$ as there are 4 different exchange paths for $J_4$ between each pair of second-neighbor tetrahedra as illustrated in Fig.~\ref{fig:Jt2}(c). 

%


\begin{thebibliography}{46}%
\makeatletter
\providecommand \@ifxundefined [1]{%
 \@ifx{#1\undefined}
}%
\providecommand \@ifnum [1]{%
 \ifnum #1\expandafter \@firstoftwo
 \else \expandafter \@secondoftwo
 \fi
}%
\providecommand \@ifx [1]{%
 \ifx #1\expandafter \@firstoftwo
 \else \expandafter \@secondoftwo
 \fi
}%
\providecommand \natexlab [1]{#1}%
\providecommand \enquote  [1]{``#1''}%
\providecommand \bibnamefont  [1]{#1}%
\providecommand \bibfnamefont [1]{#1}%
\providecommand \citenamefont [1]{#1}%
\providecommand \href@noop [0]{\@secondoftwo}%
\providecommand \href [0]{\begingroup \@sanitize@url \@href}%
\providecommand \@href[1]{\@@startlink{#1}\@@href}%
\providecommand \@@href[1]{\endgroup#1\@@endlink}%
\providecommand \@sanitize@url [0]{\catcode `\\12\catcode `\$12\catcode
  `\&12\catcode `\#12\catcode `\^12\catcode `\_12\catcode `\%12\relax}%
\providecommand \@@startlink[1]{}%
\providecommand \@@endlink[0]{}%
\providecommand \url  [0]{\begingroup\@sanitize@url \@url }%
\providecommand \@url [1]{\endgroup\@href {#1}{\urlprefix }}%
\providecommand \urlprefix  [0]{URL }%
\providecommand \Eprint [0]{\href }%
\providecommand \doibase [0]{https://doi.org/}%
\providecommand \selectlanguage [0]{\@gobble}%
\providecommand \bibinfo  [0]{\@secondoftwo}%
\providecommand \bibfield  [0]{\@secondoftwo}%
\providecommand \translation [1]{[#1]}%
\providecommand \BibitemOpen [0]{}%
\providecommand \bibitemStop [0]{}%
\providecommand \bibitemNoStop [0]{.\EOS\space}%
\providecommand \EOS [0]{\spacefactor3000\relax}%
\providecommand \BibitemShut  [1]{\csname bibitem#1\endcsname}%
\let\auto@bib@innerbib\@empty
\bibitem [{\citenamefont {Coleman}(2015)}]{coleman_intro_2016}%
  \BibitemOpen
  \bibfield  {author} {\bibinfo {author} {\bibfnamefont {P.}~\bibnamefont
  {Coleman}},\ }\bibfield  {title} {\bibinfo {title} {Renormalization
  concept},\ }in\ \href@noop {} {\emph {\bibinfo {booktitle} {Introduction to
  Many Body Physics}}}\ (\bibinfo  {publisher} {Cambridge University Press,
  Cambridge},\ \bibinfo {year} {2015})\ pp.\ \bibinfo {pages}
  {609--613}\BibitemShut {NoStop}%
\bibitem [{\citenamefont {Fazekas}(2003)}]{fazekas_2003_lecture}%
  \BibitemOpen
  \bibfield  {author} {\bibinfo {author} {\bibfnamefont {P.}~\bibnamefont
  {Fazekas}},\ }\bibfield  {title} {\bibinfo {title} {Mott insulators},\ }in\
  \href@noop {} {\emph {\bibinfo {booktitle} {Lecture Notes on Electron
  Correlation and Magnetism}}}\ (\bibinfo  {publisher} {World Scientific
  Publishing, Singapore},\ \bibinfo {year} {2003})\ pp.\ \bibinfo {pages}
  {199--214}\BibitemShut {NoStop}%
\bibitem [{\citenamefont {Castelnovo}\ \emph {et~al.}(2008)\citenamefont
  {Castelnovo}, \citenamefont {Moessner},\ and\ \citenamefont
  {Sondhi}}]{castelnovo_magnetic_2008}%
  \BibitemOpen
  \bibfield  {author} {\bibinfo {author} {\bibfnamefont {C.}~\bibnamefont
  {Castelnovo}}, \bibinfo {author} {\bibfnamefont {R.}~\bibnamefont
  {Moessner}},\ and\ \bibinfo {author} {\bibfnamefont {S.~L.}\ \bibnamefont
  {Sondhi}},\ }\bibfield  {title} {\bibinfo {title} {Magnetic monopoles in spin
  ice},\ }\href {https://doi.org/10.1038/nature06433} {\bibfield  {journal}
  {\bibinfo  {journal} {Nature}\ }\textbf {\bibinfo {volume} {451}},\ \bibinfo
  {pages} {42} (\bibinfo {year} {2008})}\BibitemShut {NoStop}%
\bibitem [{\citenamefont {Fennell}\ \emph {et~al.}(2009)\citenamefont
  {Fennell}, \citenamefont {Deen}, \citenamefont {Wildes}, \citenamefont
  {Schmalzl}, \citenamefont {Prabhakaran}, \citenamefont {Boothroyd},
  \citenamefont {Aldus}, \citenamefont {McMorrow},\ and\ \citenamefont
  {Bramwell}}]{fennell_magnetic_2009}%
  \BibitemOpen
  \bibfield  {author} {\bibinfo {author} {\bibfnamefont {T.}~\bibnamefont
  {Fennell}}, \bibinfo {author} {\bibfnamefont {P.~P.}\ \bibnamefont {Deen}},
  \bibinfo {author} {\bibfnamefont {A.~R.}\ \bibnamefont {Wildes}}, \bibinfo
  {author} {\bibfnamefont {K.}~\bibnamefont {Schmalzl}}, \bibinfo {author}
  {\bibfnamefont {D.}~\bibnamefont {Prabhakaran}}, \bibinfo {author}
  {\bibfnamefont {A.~T.}\ \bibnamefont {Boothroyd}}, \bibinfo {author}
  {\bibfnamefont {R.~J.}\ \bibnamefont {Aldus}}, \bibinfo {author}
  {\bibfnamefont {D.~F.}\ \bibnamefont {McMorrow}},\ and\ \bibinfo {author}
  {\bibfnamefont {S.~T.}\ \bibnamefont {Bramwell}},\ }\bibfield  {title}
  {\bibinfo {title} {Magnetic {Coulomb} phase in the spin ice
  {Ho$_2$Ti$_2$O$_7$}},\ }\href {https://doi.org/10.1126/science.1177582}
  {\bibfield  {journal} {\bibinfo  {journal} {Science}\ }\textbf {\bibinfo
  {volume} {326}},\ \bibinfo {pages} {415} (\bibinfo {year}
  {2009})}\BibitemShut {NoStop}%
\bibitem [{\citenamefont {Gao}\ \emph {et~al.}(2018{\natexlab{a}})\citenamefont
  {Gao}, \citenamefont {Zaharko}, \citenamefont {Tsurkan}, \citenamefont
  {Prodan}, \citenamefont {Riordan}, \citenamefont {Lago}, \citenamefont
  {F{\aa}k}, \citenamefont {Wildes}, \citenamefont {Koza}, \citenamefont
  {Ritter}, \citenamefont {Fouquet}, \citenamefont {Keller}, \citenamefont
  {Can\'evet}, \citenamefont {Medarde}, \citenamefont {Blomgren}, \citenamefont
  {Johansson}, \citenamefont {Giblin}, \citenamefont {Vrtnik}, \citenamefont
  {Luzar}, \citenamefont {Loidl}, \citenamefont {R\"uegg},\ and\ \citenamefont
  {Fennell}}]{gao_dipolar_2018}%
  \BibitemOpen
  \bibfield  {author} {\bibinfo {author} {\bibfnamefont {S.}~\bibnamefont
  {Gao}}, \bibinfo {author} {\bibfnamefont {O.}~\bibnamefont {Zaharko}},
  \bibinfo {author} {\bibfnamefont {V.}~\bibnamefont {Tsurkan}}, \bibinfo
  {author} {\bibfnamefont {L.}~\bibnamefont {Prodan}}, \bibinfo {author}
  {\bibfnamefont {E.}~\bibnamefont {Riordan}}, \bibinfo {author} {\bibfnamefont
  {J.}~\bibnamefont {Lago}}, \bibinfo {author} {\bibfnamefont {B.}~\bibnamefont
  {F{\aa}k}}, \bibinfo {author} {\bibfnamefont {A.~R.}\ \bibnamefont {Wildes}},
  \bibinfo {author} {\bibfnamefont {M.~M.}\ \bibnamefont {Koza}}, \bibinfo
  {author} {\bibfnamefont {C.}~\bibnamefont {Ritter}}, \bibinfo {author}
  {\bibfnamefont {P.}~\bibnamefont {Fouquet}}, \bibinfo {author} {\bibfnamefont
  {L.}~\bibnamefont {Keller}}, \bibinfo {author} {\bibfnamefont
  {E.}~\bibnamefont {Can\'evet}}, \bibinfo {author} {\bibfnamefont
  {M.}~\bibnamefont {Medarde}}, \bibinfo {author} {\bibfnamefont
  {J.}~\bibnamefont {Blomgren}}, \bibinfo {author} {\bibfnamefont
  {C.}~\bibnamefont {Johansson}}, \bibinfo {author} {\bibfnamefont {S.~R.}\
  \bibnamefont {Giblin}}, \bibinfo {author} {\bibfnamefont {S.}~\bibnamefont
  {Vrtnik}}, \bibinfo {author} {\bibfnamefont {J.}~\bibnamefont {Luzar}},
  \bibinfo {author} {\bibfnamefont {A.}~\bibnamefont {Loidl}}, \bibinfo
  {author} {\bibfnamefont {C.}~\bibnamefont {R\"uegg}},\ and\ \bibinfo {author}
  {\bibfnamefont {T.}~\bibnamefont {Fennell}},\ }\bibfield  {title} {\bibinfo
  {title} {Dipolar spin ice states with a fast monopole hopping rate in
  {CdEr$_2X_4$} (${X}=${Se}, {S})},\ }\href
  {https://doi.org/10.1103/PhysRevLett.120.137201} {\bibfield  {journal}
  {\bibinfo  {journal} {Phys. Rev. Lett.}\ }\textbf {\bibinfo {volume} {120}},\
  \bibinfo {pages} {137201} (\bibinfo {year} {2018}{\natexlab{a}})}\BibitemShut
  {NoStop}%
\bibitem [{\citenamefont {Gao}\ \emph {et~al.}(2019)\citenamefont {Gao},
  \citenamefont {Hirschberger}, \citenamefont {Zaharko}, \citenamefont
  {Nakajima}, \citenamefont {Kurumaji}, \citenamefont {Kikkawa}, \citenamefont
  {Shiogai}, \citenamefont {Tsukazaki}, \citenamefont {Kimura}, \citenamefont
  {Awaji}, \citenamefont {Taguchi}, \citenamefont {Arima},\ and\ \citenamefont
  {Tokura}}]{gao_ordering_2019}%
  \BibitemOpen
  \bibfield  {author} {\bibinfo {author} {\bibfnamefont {S.}~\bibnamefont
  {Gao}}, \bibinfo {author} {\bibfnamefont {M.}~\bibnamefont {Hirschberger}},
  \bibinfo {author} {\bibfnamefont {O.}~\bibnamefont {Zaharko}}, \bibinfo
  {author} {\bibfnamefont {T.}~\bibnamefont {Nakajima}}, \bibinfo {author}
  {\bibfnamefont {T.}~\bibnamefont {Kurumaji}}, \bibinfo {author}
  {\bibfnamefont {A.}~\bibnamefont {Kikkawa}}, \bibinfo {author} {\bibfnamefont
  {J.}~\bibnamefont {Shiogai}}, \bibinfo {author} {\bibfnamefont
  {A.}~\bibnamefont {Tsukazaki}}, \bibinfo {author} {\bibfnamefont
  {S.}~\bibnamefont {Kimura}}, \bibinfo {author} {\bibfnamefont
  {S.}~\bibnamefont {Awaji}}, \bibinfo {author} {\bibfnamefont
  {Y.}~\bibnamefont {Taguchi}}, \bibinfo {author} {\bibfnamefont {T.-h.}\
  \bibnamefont {Arima}},\ and\ \bibinfo {author} {\bibfnamefont
  {Y.}~\bibnamefont {Tokura}},\ }\bibfield  {title} {\bibinfo {title} {Ordering
  phenomena of spin trimers accompanied by a large geometrical {Hall} effect},\
  }\href {https://doi.org/10.1103/PhysRevB.100.241115} {\bibfield  {journal}
  {\bibinfo  {journal} {Phys. Rev. B}\ }\textbf {\bibinfo {volume} {100}},\
  \bibinfo {pages} {241115(R)} (\bibinfo {year} {2019})}\BibitemShut {NoStop}%
\bibitem [{\citenamefont {Kimura}\ \emph {et~al.}(2014)\citenamefont {Kimura},
  \citenamefont {Nakatsuji},\ and\ \citenamefont
  {Kimura}}]{kimura_experimental_2014}%
  \BibitemOpen
  \bibfield  {author} {\bibinfo {author} {\bibfnamefont {K.}~\bibnamefont
  {Kimura}}, \bibinfo {author} {\bibfnamefont {S.}~\bibnamefont {Nakatsuji}},\
  and\ \bibinfo {author} {\bibfnamefont {T.}~\bibnamefont {Kimura}},\
  }\bibfield  {title} {\bibinfo {title} {Experimental realization of a quantum
  breathing pyrochlore antiferromagnet},\ }\href
  {https://doi.org/10.1103/PhysRevB.90.060414} {\bibfield  {journal} {\bibinfo
  {journal} {Phys. Rev. B}\ }\textbf {\bibinfo {volume} {90}},\ \bibinfo
  {pages} {060414(R)} (\bibinfo {year} {2014})}\BibitemShut {NoStop}%
\bibitem [{\citenamefont {Rau}\ \emph {et~al.}(2016)\citenamefont {Rau},
  \citenamefont {Wu}, \citenamefont {May}, \citenamefont {Poudel},
  \citenamefont {Winn}, \citenamefont {Garlea}, \citenamefont {Huq},
  \citenamefont {Whitfield}, \citenamefont {Taylor}, \citenamefont {Lumsden},
  \citenamefont {Gingras},\ and\ \citenamefont
  {Christianson}}]{rau_anisotropic_2016}%
  \BibitemOpen
  \bibfield  {author} {\bibinfo {author} {\bibfnamefont {J.~G.}\ \bibnamefont
  {Rau}}, \bibinfo {author} {\bibfnamefont {L.~S.}\ \bibnamefont {Wu}},
  \bibinfo {author} {\bibfnamefont {A.~F.}\ \bibnamefont {May}}, \bibinfo
  {author} {\bibfnamefont {L.}~\bibnamefont {Poudel}}, \bibinfo {author}
  {\bibfnamefont {B.}~\bibnamefont {Winn}}, \bibinfo {author} {\bibfnamefont
  {V.~O.}\ \bibnamefont {Garlea}}, \bibinfo {author} {\bibfnamefont
  {A.}~\bibnamefont {Huq}}, \bibinfo {author} {\bibfnamefont {P.}~\bibnamefont
  {Whitfield}}, \bibinfo {author} {\bibfnamefont {A.~E.}\ \bibnamefont
  {Taylor}}, \bibinfo {author} {\bibfnamefont {M.~D.}\ \bibnamefont {Lumsden}},
  \bibinfo {author} {\bibfnamefont {M.~J.~P.}\ \bibnamefont {Gingras}},\ and\
  \bibinfo {author} {\bibfnamefont {A.~D.}\ \bibnamefont {Christianson}},\
  }\bibfield  {title} {\bibinfo {title} {Anisotropic exchange within decoupled
  tetrahedra in the quantum breathing pyrochlore
  {Ba$_3$Yb$_2$Zn$_5$O$_{11}$}},\ }\href
  {https://doi.org/10.1103/PhysRevLett.116.257204} {\bibfield  {journal}
  {\bibinfo  {journal} {Phys. Rev. Lett.}\ }\textbf {\bibinfo {volume} {116}},\
  \bibinfo {pages} {257204} (\bibinfo {year} {2016})}\BibitemShut {NoStop}%
\bibitem [{\citenamefont {Haku}\ \emph {et~al.}(2016)\citenamefont {Haku},
  \citenamefont {Kimura}, \citenamefont {Matsumoto}, \citenamefont {Soda},
  \citenamefont {Sera}, \citenamefont {Yu}, \citenamefont {Mole}, \citenamefont
  {Takeuchi}, \citenamefont {Nakatsuji}, \citenamefont {Kono}, \citenamefont
  {Sakakibara}, \citenamefont {Chang},\ and\ \citenamefont
  {Masuda}}]{haku_low_2016}%
  \BibitemOpen
  \bibfield  {author} {\bibinfo {author} {\bibfnamefont {T.}~\bibnamefont
  {Haku}}, \bibinfo {author} {\bibfnamefont {K.}~\bibnamefont {Kimura}},
  \bibinfo {author} {\bibfnamefont {Y.}~\bibnamefont {Matsumoto}}, \bibinfo
  {author} {\bibfnamefont {M.}~\bibnamefont {Soda}}, \bibinfo {author}
  {\bibfnamefont {M.}~\bibnamefont {Sera}}, \bibinfo {author} {\bibfnamefont
  {D.}~\bibnamefont {Yu}}, \bibinfo {author} {\bibfnamefont {R.~A.}\
  \bibnamefont {Mole}}, \bibinfo {author} {\bibfnamefont {T.}~\bibnamefont
  {Takeuchi}}, \bibinfo {author} {\bibfnamefont {S.}~\bibnamefont {Nakatsuji}},
  \bibinfo {author} {\bibfnamefont {Y.}~\bibnamefont {Kono}}, \bibinfo {author}
  {\bibfnamefont {T.}~\bibnamefont {Sakakibara}}, \bibinfo {author}
  {\bibfnamefont {L.-J.}\ \bibnamefont {Chang}},\ and\ \bibinfo {author}
  {\bibfnamefont {T.}~\bibnamefont {Masuda}},\ }\bibfield  {title} {\bibinfo
  {title} {Low-energy excitations and ground-state selection in the quantum
  breathing pyrochlore antiferromagnet {Ba$_3$Yb$_2$Zn$_5$O$_{11}$}},\ }\href
  {https://doi.org/10.1103/PhysRevB.93.220407} {\bibfield  {journal} {\bibinfo
  {journal} {Phys. Rev. B}\ }\textbf {\bibinfo {volume} {93}},\ \bibinfo
  {pages} {220407(R)} (\bibinfo {year} {2016})}\BibitemShut {NoStop}%
\bibitem [{\citenamefont {Janson}\ \emph {et~al.}(2014)\citenamefont {Janson},
  \citenamefont {Rousochatzakis}, \citenamefont {Tsirlin}, \citenamefont
  {Belesi}, \citenamefont {Leonov}, \citenamefont {Rößler}, \citenamefont
  {van~den Brink},\ and\ \citenamefont {Rosner}}]{janson_quantum_2014}%
  \BibitemOpen
  \bibfield  {author} {\bibinfo {author} {\bibfnamefont {O.}~\bibnamefont
  {Janson}}, \bibinfo {author} {\bibfnamefont {I.}~\bibnamefont
  {Rousochatzakis}}, \bibinfo {author} {\bibfnamefont {A.~A.}\ \bibnamefont
  {Tsirlin}}, \bibinfo {author} {\bibfnamefont {M.}~\bibnamefont {Belesi}},
  \bibinfo {author} {\bibfnamefont {A.~A.}\ \bibnamefont {Leonov}}, \bibinfo
  {author} {\bibfnamefont {U.~K.}\ \bibnamefont {Rößler}}, \bibinfo {author}
  {\bibfnamefont {J.}~\bibnamefont {van~den Brink}},\ and\ \bibinfo {author}
  {\bibfnamefont {H.}~\bibnamefont {Rosner}},\ }\bibfield  {title} {\bibinfo
  {title} {The quantum nature of skyrmions and half-skyrmions in
  {Cu$_2$OSeO$_3$}},\ }\href {https://doi.org/10.1038/ncomms6376} {\bibfield
  {journal} {\bibinfo  {journal} {Nat. Commun.}\ }\textbf {\bibinfo {volume}
  {5}},\ \bibinfo {pages} {5376} (\bibinfo {year} {2014})}\BibitemShut
  {NoStop}%
\bibitem [{\citenamefont {Romh\'anyi}\ \emph {et~al.}(2014)\citenamefont
  {Romh\'anyi}, \citenamefont {van~den Brink},\ and\ \citenamefont
  {Rousochatzakis}}]{romhanyi_entangled_2014}%
  \BibitemOpen
  \bibfield  {author} {\bibinfo {author} {\bibfnamefont {J.}~\bibnamefont
  {Romh\'anyi}}, \bibinfo {author} {\bibfnamefont {J.}~\bibnamefont {van~den
  Brink}},\ and\ \bibinfo {author} {\bibfnamefont {I.}~\bibnamefont
  {Rousochatzakis}},\ }\bibfield  {title} {\bibinfo {title} {Entangled
  tetrahedron ground state and excitations of the magnetoelectric skyrmion
  material {Cu$_2$OSeO$_3$}},\ }\href
  {https://doi.org/10.1103/PhysRevB.90.140404} {\bibfield  {journal} {\bibinfo
  {journal} {Phys. Rev. B}\ }\textbf {\bibinfo {volume} {90}},\ \bibinfo
  {pages} {140404(R)} (\bibinfo {year} {2014})}\BibitemShut {NoStop}%
\bibitem [{\citenamefont {Portnichenko}\ \emph {et~al.}(2016)\citenamefont
  {Portnichenko}, \citenamefont {Romhányi}, \citenamefont {Onykiienko},
  \citenamefont {Henschel}, \citenamefont {Schmidt}, \citenamefont {Cameron},
  \citenamefont {Surmach}, \citenamefont {Lim}, \citenamefont {Park},
  \citenamefont {Schneidewind}, \citenamefont {Abernathy}, \citenamefont
  {Rosner}, \citenamefont {van~den Brink},\ and\ \citenamefont
  {Inosov}}]{portnichenko_magnon_2016}%
  \BibitemOpen
  \bibfield  {author} {\bibinfo {author} {\bibfnamefont {P.~Y.}\ \bibnamefont
  {Portnichenko}}, \bibinfo {author} {\bibfnamefont {J.}~\bibnamefont
  {Romhányi}}, \bibinfo {author} {\bibfnamefont {Y.~A.}\ \bibnamefont
  {Onykiienko}}, \bibinfo {author} {\bibfnamefont {A.}~\bibnamefont
  {Henschel}}, \bibinfo {author} {\bibfnamefont {M.}~\bibnamefont {Schmidt}},
  \bibinfo {author} {\bibfnamefont {A.~S.}\ \bibnamefont {Cameron}}, \bibinfo
  {author} {\bibfnamefont {M.~A.}\ \bibnamefont {Surmach}}, \bibinfo {author}
  {\bibfnamefont {J.~A.}\ \bibnamefont {Lim}}, \bibinfo {author} {\bibfnamefont
  {J.~T.}\ \bibnamefont {Park}}, \bibinfo {author} {\bibfnamefont
  {A.}~\bibnamefont {Schneidewind}}, \bibinfo {author} {\bibfnamefont {D.~L.}\
  \bibnamefont {Abernathy}}, \bibinfo {author} {\bibfnamefont {H.}~\bibnamefont
  {Rosner}}, \bibinfo {author} {\bibfnamefont {J.}~\bibnamefont {van~den
  Brink}},\ and\ \bibinfo {author} {\bibfnamefont {D.~S.}\ \bibnamefont
  {Inosov}},\ }\bibfield  {title} {\bibinfo {title} {Magnon spectrum of the
  helimagnetic insulator {Cu$_2$OSeO$_3$}},\ }\href
  {https://doi.org/10.1038/ncomms10725} {\bibfield  {journal} {\bibinfo
  {journal} {Nat. Commun.}\ }\textbf {\bibinfo {volume} {7}},\ \bibinfo {pages}
  {10725} (\bibinfo {year} {2016})}\BibitemShut {NoStop}%
\bibitem [{\citenamefont {Tucker}\ \emph {et~al.}(2016)\citenamefont {Tucker},
  \citenamefont {White}, \citenamefont {Romh\'anyi}, \citenamefont {Szaller},
  \citenamefont {K\'ezsm\'arki}, \citenamefont {Roessli}, \citenamefont
  {Stuhr}, \citenamefont {Magrez}, \citenamefont {Groitl}, \citenamefont
  {Babkevich}, \citenamefont {Huang}, \citenamefont {{\v Z}ivkovi\'c},\ and\
  \citenamefont {R{\o}nnow}}]{tucker_spin_2016}%
  \BibitemOpen
  \bibfield  {author} {\bibinfo {author} {\bibfnamefont {G.~S.}\ \bibnamefont
  {Tucker}}, \bibinfo {author} {\bibfnamefont {J.~S.}\ \bibnamefont {White}},
  \bibinfo {author} {\bibfnamefont {J.}~\bibnamefont {Romh\'anyi}}, \bibinfo
  {author} {\bibfnamefont {D.}~\bibnamefont {Szaller}}, \bibinfo {author}
  {\bibfnamefont {I.}~\bibnamefont {K\'ezsm\'arki}}, \bibinfo {author}
  {\bibfnamefont {B.}~\bibnamefont {Roessli}}, \bibinfo {author} {\bibfnamefont
  {U.}~\bibnamefont {Stuhr}}, \bibinfo {author} {\bibfnamefont
  {A.}~\bibnamefont {Magrez}}, \bibinfo {author} {\bibfnamefont
  {F.}~\bibnamefont {Groitl}}, \bibinfo {author} {\bibfnamefont
  {P.}~\bibnamefont {Babkevich}}, \bibinfo {author} {\bibfnamefont
  {P.}~\bibnamefont {Huang}}, \bibinfo {author} {\bibfnamefont
  {I.}~\bibnamefont {{\v Z}ivkovi\'c}},\ and\ \bibinfo {author} {\bibfnamefont
  {H.~M.}\ \bibnamefont {R{\o}nnow}},\ }\bibfield  {title} {\bibinfo {title}
  {Spin excitations in the skyrmion host {Cu$_2$OSeO$_3$}},\ }\href
  {http://link.aps.org/doi/10.1103/PhysRevB.93.054401} {\bibfield  {journal}
  {\bibinfo  {journal} {Phys. Rev. B}\ }\textbf {\bibinfo {volume} {93}},\
  \bibinfo {pages} {054401} (\bibinfo {year} {2016})}\BibitemShut {NoStop}%
\bibitem [{\citenamefont {Pinch}\ \emph {et~al.}(1970)\citenamefont {Pinch},
  \citenamefont {Woods},\ and\ \citenamefont {Lopatin}}]{pinch_new_1970}%
  \BibitemOpen
  \bibfield  {author} {\bibinfo {author} {\bibfnamefont {H.}~\bibnamefont
  {Pinch}}, \bibinfo {author} {\bibfnamefont {M.}~\bibnamefont {Woods}},\ and\
  \bibinfo {author} {\bibfnamefont {E.}~\bibnamefont {Lopatin}},\ }\bibfield
  {title} {\bibinfo {title} {Some new mixed {A}-site chromium chalcogenide
  spinels},\ }\href {https://doi.org/10.1016/0025-5408(70)90081-4} {\bibfield
  {journal} {\bibinfo  {journal} {Mater. Res. Bull.}\ }\textbf {\bibinfo
  {volume} {5}},\ \bibinfo {pages} {425} (\bibinfo {year} {1970})}\BibitemShut
  {NoStop}%
\bibitem [{\citenamefont {van Gorkom}\ \emph {et~al.}(1973)\citenamefont {van
  Gorkom}, \citenamefont {Henning},\ and\ \citenamefont {van
  Stapele}}]{gorkom_optical_1973}%
  \BibitemOpen
  \bibfield  {author} {\bibinfo {author} {\bibfnamefont {G.~G.~P.}\
  \bibnamefont {van Gorkom}}, \bibinfo {author} {\bibfnamefont {J.~C.~M.}\
  \bibnamefont {Henning}},\ and\ \bibinfo {author} {\bibfnamefont {R.~P.}\
  \bibnamefont {van Stapele}},\ }\bibfield  {title} {\bibinfo {title} {Optical
  spectra of {Cr}$^{3+}$ pairs in the spinel {ZnGa$_2$O$_4$}},\ }\href
  {https://doi.org/10.1103/PhysRevB.8.955} {\bibfield  {journal} {\bibinfo
  {journal} {Phys. Rev. B}\ }\textbf {\bibinfo {volume} {8}},\ \bibinfo {pages}
  {955} (\bibinfo {year} {1973})}\BibitemShut {NoStop}%
\bibitem [{\citenamefont {Lee}\ \emph {et~al.}(2000)\citenamefont {Lee},
  \citenamefont {Broholm}, \citenamefont {Kim}, \citenamefont {Ratcliff~II},\
  and\ \citenamefont {Cheong}}]{lee_local_2000}%
  \BibitemOpen
  \bibfield  {author} {\bibinfo {author} {\bibfnamefont {S.~H.}\ \bibnamefont
  {Lee}}, \bibinfo {author} {\bibfnamefont {C.}~\bibnamefont {Broholm}},
  \bibinfo {author} {\bibfnamefont {T.~H.}\ \bibnamefont {Kim}}, \bibinfo
  {author} {\bibfnamefont {W.}~\bibnamefont {Ratcliff~II}},\ and\ \bibinfo
  {author} {\bibfnamefont {S.~W.}\ \bibnamefont {Cheong}},\ }\bibfield  {title}
  {\bibinfo {title} {Local spin resonance and spin-{Peierls}-like phase
  transition in a geometrically frustrated antiferromagnet},\ }\href
  {http://link.aps.org/doi/10.1103/PhysRevLett.84.3718} {\bibfield  {journal}
  {\bibinfo  {journal} {Phys. Rev. Lett.}\ }\textbf {\bibinfo {volume} {84}},\
  \bibinfo {pages} {3718} (\bibinfo {year} {2000})}\BibitemShut {NoStop}%
\bibitem [{\citenamefont {Okamoto}\ \emph {et~al.}(2013)\citenamefont
  {Okamoto}, \citenamefont {Nilsen}, \citenamefont {Attfield},\ and\
  \citenamefont {Hiroi}}]{okamoto_breathing_2013}%
  \BibitemOpen
  \bibfield  {author} {\bibinfo {author} {\bibfnamefont {Y.}~\bibnamefont
  {Okamoto}}, \bibinfo {author} {\bibfnamefont {G.~J.}\ \bibnamefont {Nilsen}},
  \bibinfo {author} {\bibfnamefont {J.~P.}\ \bibnamefont {Attfield}},\ and\
  \bibinfo {author} {\bibfnamefont {Z.}~\bibnamefont {Hiroi}},\ }\bibfield
  {title} {\bibinfo {title} {Breathing pyrochlore lattice realized in
  {$A$}-site ordered spinel oxides {LiGaCr$_4$O$_8$} and {LiInCr$_4$O$_8$}},\
  }\href {http://link.aps.org/doi/10.1103/PhysRevLett.110.097203} {\bibfield
  {journal} {\bibinfo  {journal} {Phys. Rev. Lett.}\ }\textbf {\bibinfo
  {volume} {110}},\ \bibinfo {pages} {97203} (\bibinfo {year}
  {2013})}\BibitemShut {NoStop}%
\bibitem [{\citenamefont {Ghosh}\ \emph {et~al.}(2019)\citenamefont {Ghosh},
  \citenamefont {Iqbal}, \citenamefont {M\"uller}, \citenamefont {Ponnaganti},
  \citenamefont {Thomale}, \citenamefont {Narayanan}, \citenamefont {Reuther},
  \citenamefont {Gingras},\ and\ \citenamefont
  {Jeschke}}]{ghosh_breathing_2019}%
  \BibitemOpen
  \bibfield  {author} {\bibinfo {author} {\bibfnamefont {P.}~\bibnamefont
  {Ghosh}}, \bibinfo {author} {\bibfnamefont {Y.}~\bibnamefont {Iqbal}},
  \bibinfo {author} {\bibfnamefont {T.}~\bibnamefont {M\"uller}}, \bibinfo
  {author} {\bibfnamefont {R.~T.}\ \bibnamefont {Ponnaganti}}, \bibinfo
  {author} {\bibfnamefont {R.}~\bibnamefont {Thomale}}, \bibinfo {author}
  {\bibfnamefont {R.}~\bibnamefont {Narayanan}}, \bibinfo {author}
  {\bibfnamefont {J.}~\bibnamefont {Reuther}}, \bibinfo {author} {\bibfnamefont
  {M.~J.~P.}\ \bibnamefont {Gingras}},\ and\ \bibinfo {author} {\bibfnamefont
  {H.~O.}\ \bibnamefont {Jeschke}},\ }\bibfield  {title} {\bibinfo {title}
  {Breathing chromium spinels: a showcase for a variety of pyrochlore
  {Heisenberg} {Hamiltonians}},\ }\href
  {https://doi.org/10.1038/s41535-019-0202-z} {\bibfield  {journal} {\bibinfo
  {journal} {Npj Quantum Mater.}\ }\textbf {\bibinfo {volume} {4}},\ \bibinfo
  {pages} {63} (\bibinfo {year} {2019})}\BibitemShut {NoStop}%
\bibitem [{\citenamefont {Pokharel}\ \emph {et~al.}(2018)\citenamefont
  {Pokharel}, \citenamefont {May}, \citenamefont {Parker}, \citenamefont
  {Calder}, \citenamefont {Ehlers}, \citenamefont {Huq}, \citenamefont
  {Kimber}, \citenamefont {Arachchige}, \citenamefont {Poudel}, \citenamefont
  {McGuire}, \citenamefont {Mandrus},\ and\ \citenamefont
  {Christianson}}]{pokharel_negative_2018}%
  \BibitemOpen
  \bibfield  {author} {\bibinfo {author} {\bibfnamefont {G.}~\bibnamefont
  {Pokharel}}, \bibinfo {author} {\bibfnamefont {A.~F.}\ \bibnamefont {May}},
  \bibinfo {author} {\bibfnamefont {D.~S.}\ \bibnamefont {Parker}}, \bibinfo
  {author} {\bibfnamefont {S.}~\bibnamefont {Calder}}, \bibinfo {author}
  {\bibfnamefont {G.}~\bibnamefont {Ehlers}}, \bibinfo {author} {\bibfnamefont
  {A.}~\bibnamefont {Huq}}, \bibinfo {author} {\bibfnamefont {S.~A.~J.}\
  \bibnamefont {Kimber}}, \bibinfo {author} {\bibfnamefont {H.~S.}\
  \bibnamefont {Arachchige}}, \bibinfo {author} {\bibfnamefont
  {L.}~\bibnamefont {Poudel}}, \bibinfo {author} {\bibfnamefont {M.~A.}\
  \bibnamefont {McGuire}}, \bibinfo {author} {\bibfnamefont {D.}~\bibnamefont
  {Mandrus}},\ and\ \bibinfo {author} {\bibfnamefont {A.~D.}\ \bibnamefont
  {Christianson}},\ }\bibfield  {title} {\bibinfo {title} {Negative thermal
  expansion and magnetoelastic coupling in the breathing pyrochlore lattice
  material {LiGaCr$_{4}$S$_{8}$}},\ }\href
  {https://doi.org/10.1103/PhysRevB.97.134117} {\bibfield  {journal} {\bibinfo
  {journal} {Phys. Rev. B}\ }\textbf {\bibinfo {volume} {97}},\ \bibinfo
  {pages} {134117} (\bibinfo {year} {2018})}\BibitemShut {NoStop}%
\bibitem [{\citenamefont {Pokharel}\ \emph {et~al.}(2020)\citenamefont
  {Pokharel}, \citenamefont {Arachchige}, \citenamefont {Williams},
  \citenamefont {May}, \citenamefont {Fishman}, \citenamefont {Sala},
  \citenamefont {Calder}, \citenamefont {Ehlers}, \citenamefont {Parker},
  \citenamefont {Hong}, \citenamefont {Wildes}, \citenamefont {Mandrus},
  \citenamefont {Paddison},\ and\ \citenamefont
  {Christianson}}]{pokharel_cluster_2020}%
  \BibitemOpen
  \bibfield  {author} {\bibinfo {author} {\bibfnamefont {G.}~\bibnamefont
  {Pokharel}}, \bibinfo {author} {\bibfnamefont {H.~S.}\ \bibnamefont
  {Arachchige}}, \bibinfo {author} {\bibfnamefont {T.~J.}\ \bibnamefont
  {Williams}}, \bibinfo {author} {\bibfnamefont {A.~F.}\ \bibnamefont {May}},
  \bibinfo {author} {\bibfnamefont {R.~S.}\ \bibnamefont {Fishman}}, \bibinfo
  {author} {\bibfnamefont {G.}~\bibnamefont {Sala}}, \bibinfo {author}
  {\bibfnamefont {S.}~\bibnamefont {Calder}}, \bibinfo {author} {\bibfnamefont
  {G.}~\bibnamefont {Ehlers}}, \bibinfo {author} {\bibfnamefont {D.~S.}\
  \bibnamefont {Parker}}, \bibinfo {author} {\bibfnamefont {T.}~\bibnamefont
  {Hong}}, \bibinfo {author} {\bibfnamefont {A.}~\bibnamefont {Wildes}},
  \bibinfo {author} {\bibfnamefont {D.}~\bibnamefont {Mandrus}}, \bibinfo
  {author} {\bibfnamefont {J.~A.~M.}\ \bibnamefont {Paddison}},\ and\ \bibinfo
  {author} {\bibfnamefont {A.~D.}\ \bibnamefont {Christianson}},\ }\bibfield
  {title} {\bibinfo {title} {Cluster frustration in the breathing pyrochlore
  magnet {LiGaCr$_{4}$S$_{8}$}},\ }\href
  {https://doi.org/10.1103/PhysRevLett.125.167201} {\bibfield  {journal}
  {\bibinfo  {journal} {Phys. Rev. Lett.}\ }\textbf {\bibinfo {volume} {125}},\
  \bibinfo {pages} {167201} (\bibinfo {year} {2020})}\BibitemShut {NoStop}%
\bibitem [{\citenamefont {Plumier}\ \emph {et~al.}(1971)\citenamefont
  {Plumier}, \citenamefont {Lotgering},\ and\ \citenamefont {van
  Stapele}}]{plumier_magnetic_1971}%
  \BibitemOpen
  \bibfield  {author} {\bibinfo {author} {\bibfnamefont {R.}~\bibnamefont
  {Plumier}}, \bibinfo {author} {\bibfnamefont {F.~K.}\ \bibnamefont
  {Lotgering}},\ and\ \bibinfo {author} {\bibfnamefont {R.~P.}\ \bibnamefont
  {van Stapele}},\ }\bibfield  {title} {\bibinfo {title} {Magnetic properties
  of {Cu$_{1/2}$In$_{1/2}$Cr$_2$S$_4$} and some related compounds},\ }\href
  {https://doi.org/10.1051/jphyscol:19711107} {\bibfield  {journal} {\bibinfo
  {journal} {J. Phys. Colloques}\ }\textbf {\bibinfo {volume} {32}},\ \bibinfo
  {pages} {324} (\bibinfo {year} {1971})}\BibitemShut {NoStop}%
\bibitem [{\citenamefont {Okamoto}\ \emph {et~al.}(2018)\citenamefont
  {Okamoto}, \citenamefont {Mori}, \citenamefont {Katayama}, \citenamefont
  {Miyake}, \citenamefont {Tokunaga}, \citenamefont {Matsuo}, \citenamefont
  {Kindo},\ and\ \citenamefont {Takenaka}}]{okamoto_magnetic_2018}%
  \BibitemOpen
  \bibfield  {author} {\bibinfo {author} {\bibfnamefont {Y.}~\bibnamefont
  {Okamoto}}, \bibinfo {author} {\bibfnamefont {M.}~\bibnamefont {Mori}},
  \bibinfo {author} {\bibfnamefont {N.}~\bibnamefont {Katayama}}, \bibinfo
  {author} {\bibfnamefont {A.}~\bibnamefont {Miyake}}, \bibinfo {author}
  {\bibfnamefont {M.}~\bibnamefont {Tokunaga}}, \bibinfo {author}
  {\bibfnamefont {A.}~\bibnamefont {Matsuo}}, \bibinfo {author} {\bibfnamefont
  {K.}~\bibnamefont {Kindo}},\ and\ \bibinfo {author} {\bibfnamefont
  {K.}~\bibnamefont {Takenaka}},\ }\bibfield  {title} {\bibinfo {title}
  {Magnetic and structural properties of {$A$}-site ordered chromium spinel
  sulfides: {Alternating} antiferromagnetic and ferromagnetic interactions in
  the breathing pyrochlore lattice},\ }\href
  {https://doi.org/10.7566/JPSJ.87.034709} {\bibfield  {journal} {\bibinfo
  {journal} {J. Phys. Soc. Jpn.}\ }\textbf {\bibinfo {volume} {87}},\ \bibinfo
  {pages} {034709} (\bibinfo {year} {2018})}\BibitemShut {NoStop}%
\bibitem [{\citenamefont {Benton}\ and\ \citenamefont
  {Shannon}(2015)}]{benton_ground_2015}%
  \BibitemOpen
  \bibfield  {author} {\bibinfo {author} {\bibfnamefont {O.}~\bibnamefont
  {Benton}}\ and\ \bibinfo {author} {\bibfnamefont {N.}~\bibnamefont
  {Shannon}},\ }\bibfield  {title} {\bibinfo {title} {Ground state selection
  and spin-liquid behaviour in the classical {Heisenberg} model on the
  breathing pyrochlore lattice},\ }\href
  {https://doi.org/10.7566/JPSJ.84.104710} {\bibfield  {journal} {\bibinfo
  {journal} {J. Phys. Soc. Jpn.}\ }\textbf {\bibinfo {volume} {84}},\ \bibinfo
  {pages} {104710} (\bibinfo {year} {2015})}\BibitemShut {NoStop}%
\bibitem [{\citenamefont {Gao}\ \emph {et~al.}(2018{\natexlab{b}})\citenamefont
  {Gao}, \citenamefont {Guratinder}, \citenamefont {Stuhr}, \citenamefont
  {White}, \citenamefont {Mansson}, \citenamefont {Roessli}, \citenamefont
  {Fennell}, \citenamefont {Tsurkan}, \citenamefont {Loidl}, \citenamefont
  {Ciomaga~Hatnean}, \citenamefont {Balakrishnan}, \citenamefont {Raymond},
  \citenamefont {Chapon}, \citenamefont {Garlea}, \citenamefont {Savici},
  \citenamefont {Cervellino}, \citenamefont {Bombardi}, \citenamefont
  {Chernyshov}, \citenamefont {R\"uegg}, \citenamefont {Haraldsen},\ and\
  \citenamefont {Zaharko}}]{gao_manifolds_2018}%
  \BibitemOpen
  \bibfield  {author} {\bibinfo {author} {\bibfnamefont {S.}~\bibnamefont
  {Gao}}, \bibinfo {author} {\bibfnamefont {K.}~\bibnamefont {Guratinder}},
  \bibinfo {author} {\bibfnamefont {U.}~\bibnamefont {Stuhr}}, \bibinfo
  {author} {\bibfnamefont {J.~S.}\ \bibnamefont {White}}, \bibinfo {author}
  {\bibfnamefont {M.}~\bibnamefont {Mansson}}, \bibinfo {author} {\bibfnamefont
  {B.}~\bibnamefont {Roessli}}, \bibinfo {author} {\bibfnamefont
  {T.}~\bibnamefont {Fennell}}, \bibinfo {author} {\bibfnamefont
  {V.}~\bibnamefont {Tsurkan}}, \bibinfo {author} {\bibfnamefont
  {A.}~\bibnamefont {Loidl}}, \bibinfo {author} {\bibfnamefont
  {M.}~\bibnamefont {Ciomaga~Hatnean}}, \bibinfo {author} {\bibfnamefont
  {G.}~\bibnamefont {Balakrishnan}}, \bibinfo {author} {\bibfnamefont
  {S.}~\bibnamefont {Raymond}}, \bibinfo {author} {\bibfnamefont
  {L.}~\bibnamefont {Chapon}}, \bibinfo {author} {\bibfnamefont {V.~O.}\
  \bibnamefont {Garlea}}, \bibinfo {author} {\bibfnamefont {A.~T.}\
  \bibnamefont {Savici}}, \bibinfo {author} {\bibfnamefont {A.}~\bibnamefont
  {Cervellino}}, \bibinfo {author} {\bibfnamefont {A.}~\bibnamefont
  {Bombardi}}, \bibinfo {author} {\bibfnamefont {D.}~\bibnamefont
  {Chernyshov}}, \bibinfo {author} {\bibfnamefont {C.}~\bibnamefont {R\"uegg}},
  \bibinfo {author} {\bibfnamefont {J.~T.}\ \bibnamefont {Haraldsen}},\ and\
  \bibinfo {author} {\bibfnamefont {O.}~\bibnamefont {Zaharko}},\ }\bibfield
  {title} {\bibinfo {title} {Manifolds of magnetic ordered states and
  excitations in the almost {Heisenberg} pyrochlore antiferromagnet
  {MgCr$_2$O$_4$}},\ }\href {https://doi.org/10.1103/PhysRevB.97.134430}
  {\bibfield  {journal} {\bibinfo  {journal} {Phys. Rev. B}\ }\textbf {\bibinfo
  {volume} {97}},\ \bibinfo {pages} {134430} (\bibinfo {year}
  {2018}{\natexlab{b}})}\BibitemShut {NoStop}%
\bibitem [{\citenamefont {Gen}\ \emph {et~al.}(2020)\citenamefont {Gen},
  \citenamefont {Okamoto}, \citenamefont {Mori}, \citenamefont {Takenaka},\
  and\ \citenamefont {Kohama}}]{gen_magnetization_2020}%
  \BibitemOpen
  \bibfield  {author} {\bibinfo {author} {\bibfnamefont {M.}~\bibnamefont
  {Gen}}, \bibinfo {author} {\bibfnamefont {Y.}~\bibnamefont {Okamoto}},
  \bibinfo {author} {\bibfnamefont {M.}~\bibnamefont {Mori}}, \bibinfo {author}
  {\bibfnamefont {K.}~\bibnamefont {Takenaka}},\ and\ \bibinfo {author}
  {\bibfnamefont {Y.}~\bibnamefont {Kohama}},\ }\bibfield  {title} {\bibinfo
  {title} {Magnetization process of the breathing pyrochlore magnet
  {CuInCr$_4$S$_8$} in ultrahigh magnetic fields up to 150 {T}},\ }\href
  {https://doi.org/10.1103/PhysRevB.101.054434} {\bibfield  {journal} {\bibinfo
   {journal} {Phys. Rev. B}\ }\textbf {\bibinfo {volume} {101}},\ \bibinfo
  {pages} {054434} (\bibinfo {year} {2020})}\BibitemShut {NoStop}%
\bibitem [{\citenamefont {Penc}\ \emph {et~al.}(2004)\citenamefont {Penc},
  \citenamefont {Shannon},\ and\ \citenamefont {Shiba}}]{penc_half_2004}%
  \BibitemOpen
  \bibfield  {author} {\bibinfo {author} {\bibfnamefont {K.}~\bibnamefont
  {Penc}}, \bibinfo {author} {\bibfnamefont {N.}~\bibnamefont {Shannon}},\ and\
  \bibinfo {author} {\bibfnamefont {H.}~\bibnamefont {Shiba}},\ }\bibfield
  {title} {\bibinfo {title} {Half-magnetization plateau stabilized by
  structural distortion in the antiferromagnetic {Heisenberg} model on a
  pyrochlore lattice},\ }\href
  {http://link.aps.org/doi/10.1103/PhysRevLett.93.197203} {\bibfield  {journal}
  {\bibinfo  {journal} {Phys. Rev. Lett.}\ }\textbf {\bibinfo {volume} {93}},\
  \bibinfo {pages} {197203} (\bibinfo {year} {2004})}\BibitemShut {NoStop}%
\bibitem [{\citenamefont {Bergman}\ \emph {et~al.}(2006)\citenamefont
  {Bergman}, \citenamefont {Shindou}, \citenamefont {Fiete},\ and\
  \citenamefont {Balents}}]{bergman_models_2006}%
  \BibitemOpen
  \bibfield  {author} {\bibinfo {author} {\bibfnamefont {D.~L.}\ \bibnamefont
  {Bergman}}, \bibinfo {author} {\bibfnamefont {R.}~\bibnamefont {Shindou}},
  \bibinfo {author} {\bibfnamefont {G.~A.}\ \bibnamefont {Fiete}},\ and\
  \bibinfo {author} {\bibfnamefont {L.}~\bibnamefont {Balents}},\ }\bibfield
  {title} {\bibinfo {title} {Models of degeneracy breaking in pyrochlore
  antiferromagnets},\ }\href
  {http://link.aps.org/doi/10.1103/PhysRevB.74.134409} {\bibfield  {journal}
  {\bibinfo  {journal} {Phys. Rev. B}\ }\textbf {\bibinfo {volume} {74}},\
  \bibinfo {pages} {134409} (\bibinfo {year} {2006})}\BibitemShut {NoStop}%
\bibitem [{sup()}]{supp}%
  \BibitemOpen
  \href@noop {} {\ }\bibinfo {note} {See Supplemental Materials for
  experimental and theoretical methods, basic characterizations of the sample,
  representation analysis, tetrahedron excitations, and further details of the
  effective cluster model.}\BibitemShut {Stop}%
\bibitem [{\citenamefont {Oh}\ \emph {et~al.}(2016)\citenamefont {Oh},
  \citenamefont {Le}, \citenamefont {Nahm}, \citenamefont {Sim}, \citenamefont
  {Jeong}, \citenamefont {Perring}, \citenamefont {Woo}, \citenamefont
  {Nakajima}, \citenamefont {Ohira-Kawamura}, \citenamefont {Yamani},
  \citenamefont {Yoshida}, \citenamefont {Eisaki}, \citenamefont {Cheong},
  \citenamefont {Chernyshev},\ and\ \citenamefont
  {Park}}]{oh_spontaneous_2016}%
  \BibitemOpen
  \bibfield  {author} {\bibinfo {author} {\bibfnamefont {J.}~\bibnamefont
  {Oh}}, \bibinfo {author} {\bibfnamefont {M.~D.}\ \bibnamefont {Le}}, \bibinfo
  {author} {\bibfnamefont {H.-H.}\ \bibnamefont {Nahm}}, \bibinfo {author}
  {\bibfnamefont {H.}~\bibnamefont {Sim}}, \bibinfo {author} {\bibfnamefont
  {J.}~\bibnamefont {Jeong}}, \bibinfo {author} {\bibfnamefont {T.~G.}\
  \bibnamefont {Perring}}, \bibinfo {author} {\bibfnamefont {H.}~\bibnamefont
  {Woo}}, \bibinfo {author} {\bibfnamefont {K.}~\bibnamefont {Nakajima}},
  \bibinfo {author} {\bibfnamefont {S.}~\bibnamefont {Ohira-Kawamura}},
  \bibinfo {author} {\bibfnamefont {Z.}~\bibnamefont {Yamani}}, \bibinfo
  {author} {\bibfnamefont {Y.}~\bibnamefont {Yoshida}}, \bibinfo {author}
  {\bibfnamefont {H.}~\bibnamefont {Eisaki}}, \bibinfo {author} {\bibfnamefont
  {S.~W.}\ \bibnamefont {Cheong}}, \bibinfo {author} {\bibfnamefont {A.~L.}\
  \bibnamefont {Chernyshev}},\ and\ \bibinfo {author} {\bibfnamefont {J.-G.}\
  \bibnamefont {Park}},\ }\bibfield  {title} {\bibinfo {title} {Spontaneous
  decays of magneto-elastic excitations in non-collinear antiferromagnet
  ({Y},{Lu}){MnO$_3$}},\ }\href {https://doi.org/10.1038/ncomms13146}
  {\bibfield  {journal} {\bibinfo  {journal} {Nat. Commun.}\ }\textbf {\bibinfo
  {volume} {7}},\ \bibinfo {pages} {13146} (\bibinfo {year}
  {2016})}\BibitemShut {NoStop}%
\bibitem [{\citenamefont {T\'oth}\ \emph {et~al.}(2016)\citenamefont {T\'oth},
  \citenamefont {Wehinger}, \citenamefont {Rolfs}, \citenamefont {Birol},
  \citenamefont {Stuhr}, \citenamefont {Takatsu}, \citenamefont {Kimura},
  \citenamefont {Kimura}, \citenamefont {R{\o}nnow},\ and\ \citenamefont
  {R\"uegg}}]{toth_electromagnon_2016}%
  \BibitemOpen
  \bibfield  {author} {\bibinfo {author} {\bibfnamefont {S.}~\bibnamefont
  {T\'oth}}, \bibinfo {author} {\bibfnamefont {B.}~\bibnamefont {Wehinger}},
  \bibinfo {author} {\bibfnamefont {K.}~\bibnamefont {Rolfs}}, \bibinfo
  {author} {\bibfnamefont {T.}~\bibnamefont {Birol}}, \bibinfo {author}
  {\bibfnamefont {U.}~\bibnamefont {Stuhr}}, \bibinfo {author} {\bibfnamefont
  {H.}~\bibnamefont {Takatsu}}, \bibinfo {author} {\bibfnamefont
  {K.}~\bibnamefont {Kimura}}, \bibinfo {author} {\bibfnamefont
  {T.}~\bibnamefont {Kimura}}, \bibinfo {author} {\bibfnamefont {H.~M.}\
  \bibnamefont {R{\o}nnow}},\ and\ \bibinfo {author} {\bibfnamefont
  {C.}~\bibnamefont {R\"uegg}},\ }\bibfield  {title} {\bibinfo {title}
  {Electromagnon dispersion probed by inelastic {X}-ray scattering in
  {LiCrO$_2$}},\ }\href {http://dx.doi.org/10.1038/ncomms13547} {\bibfield
  {journal} {\bibinfo  {journal} {Nat. Commun.}\ }\textbf {\bibinfo {volume}
  {7}},\ \bibinfo {pages} {13547} (\bibinfo {year} {2016})}\BibitemShut
  {NoStop}%
\bibitem [{\citenamefont {Hallas}\ \emph {et~al.}(2016)\citenamefont {Hallas},
  \citenamefont {Gaudet}, \citenamefont {Butch}, \citenamefont {Tachibana},
  \citenamefont {Freitas}, \citenamefont {Luke}, \citenamefont {Wiebe},\ and\
  \citenamefont {Gaulin}}]{hallas_universal_2016}%
  \BibitemOpen
  \bibfield  {author} {\bibinfo {author} {\bibfnamefont {A.~M.}\ \bibnamefont
  {Hallas}}, \bibinfo {author} {\bibfnamefont {J.}~\bibnamefont {Gaudet}},
  \bibinfo {author} {\bibfnamefont {N.~P.}\ \bibnamefont {Butch}}, \bibinfo
  {author} {\bibfnamefont {M.}~\bibnamefont {Tachibana}}, \bibinfo {author}
  {\bibfnamefont {R.~S.}\ \bibnamefont {Freitas}}, \bibinfo {author}
  {\bibfnamefont {G.~M.}\ \bibnamefont {Luke}}, \bibinfo {author}
  {\bibfnamefont {C.~R.}\ \bibnamefont {Wiebe}},\ and\ \bibinfo {author}
  {\bibfnamefont {B.~D.}\ \bibnamefont {Gaulin}},\ }\bibfield  {title}
  {\bibinfo {title} {Universal dynamic magnetism in {Yb} pyrochlores with
  disparate ground states},\ }\href
  {https://doi.org/10.1103/PhysRevB.93.100403} {\bibfield  {journal} {\bibinfo
  {journal} {Phys. Rev. B}\ }\textbf {\bibinfo {volume} {93}},\ \bibinfo
  {pages} {100403(R)} (\bibinfo {year} {2016})}\BibitemShut {NoStop}%
\bibitem [{\citenamefont {Guratinder}\ \emph {et~al.}(2019)\citenamefont
  {Guratinder}, \citenamefont {Rau}, \citenamefont {Tsurkan}, \citenamefont
  {Ritter}, \citenamefont {Embs}, \citenamefont {Fennell}, \citenamefont
  {Walker}, \citenamefont {Medarde}, \citenamefont {Shang}, \citenamefont
  {Cervellino}, \citenamefont {R\"uegg},\ and\ \citenamefont
  {Zaharko}}]{guratinder_multi_2019}%
  \BibitemOpen
  \bibfield  {author} {\bibinfo {author} {\bibfnamefont {K.}~\bibnamefont
  {Guratinder}}, \bibinfo {author} {\bibfnamefont {J.~G.}\ \bibnamefont {Rau}},
  \bibinfo {author} {\bibfnamefont {V.}~\bibnamefont {Tsurkan}}, \bibinfo
  {author} {\bibfnamefont {C.}~\bibnamefont {Ritter}}, \bibinfo {author}
  {\bibfnamefont {J.}~\bibnamefont {Embs}}, \bibinfo {author} {\bibfnamefont
  {T.}~\bibnamefont {Fennell}}, \bibinfo {author} {\bibfnamefont {H.~C.}\
  \bibnamefont {Walker}}, \bibinfo {author} {\bibfnamefont {M.}~\bibnamefont
  {Medarde}}, \bibinfo {author} {\bibfnamefont {T.}~\bibnamefont {Shang}},
  \bibinfo {author} {\bibfnamefont {A.}~\bibnamefont {Cervellino}}, \bibinfo
  {author} {\bibfnamefont {C.}~\bibnamefont {R\"uegg}},\ and\ \bibinfo {author}
  {\bibfnamefont {O.}~\bibnamefont {Zaharko}},\ }\bibfield  {title} {\bibinfo
  {title} {Multiphase competition in the quantum {XY} pyrochlore
  antiferromagnet {CdYb$_2$Se$_4$}: Zero and applied magnetic field study},\
  }\href {https://doi.org/10.1103/PhysRevB.100.094420} {\bibfield  {journal}
  {\bibinfo  {journal} {Phys. Rev. B}\ }\textbf {\bibinfo {volume} {100}},\
  \bibinfo {pages} {094420} (\bibinfo {year} {2019})}\BibitemShut {NoStop}%
\bibitem [{\citenamefont {Tomiyasu}\ \emph {et~al.}(2008)\citenamefont
  {Tomiyasu}, \citenamefont {Suzuki}, \citenamefont {Toki}, \citenamefont
  {Itoh}, \citenamefont {Matsuura}, \citenamefont {Aso},\ and\ \citenamefont
  {Yamada}}]{tomiyasu_molecular_2008}%
  \BibitemOpen
  \bibfield  {author} {\bibinfo {author} {\bibfnamefont {K.}~\bibnamefont
  {Tomiyasu}}, \bibinfo {author} {\bibfnamefont {H.}~\bibnamefont {Suzuki}},
  \bibinfo {author} {\bibfnamefont {M.}~\bibnamefont {Toki}}, \bibinfo {author}
  {\bibfnamefont {S.}~\bibnamefont {Itoh}}, \bibinfo {author} {\bibfnamefont
  {M.}~\bibnamefont {Matsuura}}, \bibinfo {author} {\bibfnamefont
  {N.}~\bibnamefont {Aso}},\ and\ \bibinfo {author} {\bibfnamefont
  {K.}~\bibnamefont {Yamada}},\ }\bibfield  {title} {\bibinfo {title}
  {Molecular spin resonance in the geometrically frustrated magnet
  {MgCr$_2$O$_4$} by inelastic neutron scattering},\ }\href
  {http://link.aps.org/doi/10.1103/PhysRevLett.101.177401} {\bibfield
  {journal} {\bibinfo  {journal} {Phys. Rev. Lett.}\ }\textbf {\bibinfo
  {volume} {101}},\ \bibinfo {pages} {177401} (\bibinfo {year}
  {2008})}\BibitemShut {NoStop}%
\bibitem [{\citenamefont {Haraldsen}\ \emph {et~al.}(2005)\citenamefont
  {Haraldsen}, \citenamefont {Barnes},\ and\ \citenamefont
  {Musfeldt}}]{haraldsen_neutron_2005}%
  \BibitemOpen
  \bibfield  {author} {\bibinfo {author} {\bibfnamefont {J.~T.}\ \bibnamefont
  {Haraldsen}}, \bibinfo {author} {\bibfnamefont {T.}~\bibnamefont {Barnes}},\
  and\ \bibinfo {author} {\bibfnamefont {J.~L.}\ \bibnamefont {Musfeldt}},\
  }\bibfield  {title} {\bibinfo {title} {Neutron scattering and magnetic
  observables for ${S}=1/2$ spin clusters and molecular magnets},\ }\href
  {http://link.aps.org/doi/10.1103/PhysRevB.71.064403} {\bibfield  {journal}
  {\bibinfo  {journal} {Phys. Rev. B}\ }\textbf {\bibinfo {volume} {71}},\
  \bibinfo {pages} {064403} (\bibinfo {year} {2005})}\BibitemShut {NoStop}%
\bibitem [{\citenamefont {Houchins}\ and\ \citenamefont
  {Haraldsen}(2015)}]{houchins_generalization_2015}%
  \BibitemOpen
  \bibfield  {author} {\bibinfo {author} {\bibfnamefont {G.}~\bibnamefont
  {Houchins}}\ and\ \bibinfo {author} {\bibfnamefont {J.~T.}\ \bibnamefont
  {Haraldsen}},\ }\bibfield  {title} {\bibinfo {title} {Generalization of
  polarized spin excitations for asymmetric dimeric systems},\ }\href
  {https://doi.org/10.1103/PhysRevB.91.014422} {\bibfield  {journal} {\bibinfo
  {journal} {Phys. Rev. B}\ }\textbf {\bibinfo {volume} {91}},\ \bibinfo
  {pages} {014422} (\bibinfo {year} {2015})}\BibitemShut {NoStop}%
\bibitem [{\citenamefont {Toth}\ and\ \citenamefont
  {Lake}(2015)}]{toth_linear_2015}%
  \BibitemOpen
  \bibfield  {author} {\bibinfo {author} {\bibfnamefont {S.}~\bibnamefont
  {Toth}}\ and\ \bibinfo {author} {\bibfnamefont {B.}~\bibnamefont {Lake}},\
  }\bibfield  {title} {\bibinfo {title} {Linear spin wave theory for single-{Q}
  incommensurate magnetic structures},\ }\href
  {https://doi.org/10.1088/0953-8984/27/16/166002} {\bibfield  {journal}
  {\bibinfo  {journal} {J. Phys.: Condens. Matter}\ }\textbf {\bibinfo {volume}
  {27}},\ \bibinfo {pages} {166002} (\bibinfo {year} {2015})}\BibitemShut
  {NoStop}%
\bibitem [{\citenamefont {Balla}\ \emph {et~al.}(2020)\citenamefont {Balla},
  \citenamefont {Iqbal},\ and\ \citenamefont {Penc}}]{balla_degenerate_2020}%
  \BibitemOpen
  \bibfield  {author} {\bibinfo {author} {\bibfnamefont {P.}~\bibnamefont
  {Balla}}, \bibinfo {author} {\bibfnamefont {Y.}~\bibnamefont {Iqbal}},\ and\
  \bibinfo {author} {\bibfnamefont {K.}~\bibnamefont {Penc}},\ }\bibfield
  {title} {\bibinfo {title} {Degenerate manifolds, helimagnets, and multi-$q$
  chiral phases in the classical {Heisenberg} antiferromagnet on the
  face-centered-cubic lattice},\ }\href
  {https://doi.org/10.1103/PhysRevResearch.2.043278} {\bibfield  {journal}
  {\bibinfo  {journal} {Phys. Rev. Research}\ }\textbf {\bibinfo {volume}
  {2}},\ \bibinfo {pages} {043278} (\bibinfo {year} {2020})}\BibitemShut
  {NoStop}%
\bibitem [{\citenamefont {Wysocki}\ \emph {et~al.}(2011)\citenamefont
  {Wysocki}, \citenamefont {Belashchenko},\ and\ \citenamefont
  {Antropov}}]{wysocki_consistent_2011}%
  \BibitemOpen
  \bibfield  {author} {\bibinfo {author} {\bibfnamefont {A.~L.}\ \bibnamefont
  {Wysocki}}, \bibinfo {author} {\bibfnamefont {K.~D.}\ \bibnamefont
  {Belashchenko}},\ and\ \bibinfo {author} {\bibfnamefont {V.~P.}\ \bibnamefont
  {Antropov}},\ }\bibfield  {title} {\bibinfo {title} {Consistent model of
  magnetism in ferropnictides},\ }\href {https://doi.org/10.1038/nphys1933}
  {\bibfield  {journal} {\bibinfo  {journal} {Nat. Phys.}\ }\textbf {\bibinfo
  {volume} {7}},\ \bibinfo {pages} {485} (\bibinfo {year} {2011})}\BibitemShut
  {NoStop}%
\bibitem [{\citenamefont {Wildes}\ \emph {et~al.}(2020)\citenamefont {Wildes},
  \citenamefont {Zhitomirsky}, \citenamefont {Ziman}, \citenamefont {Lan{\c
  o}on},\ and\ \citenamefont {Walker}}]{wildes_evidence_2020}%
  \BibitemOpen
  \bibfield  {author} {\bibinfo {author} {\bibfnamefont {A.~R.}\ \bibnamefont
  {Wildes}}, \bibinfo {author} {\bibfnamefont {M.~E.}\ \bibnamefont
  {Zhitomirsky}}, \bibinfo {author} {\bibfnamefont {T.}~\bibnamefont {Ziman}},
  \bibinfo {author} {\bibfnamefont {D.}~\bibnamefont {Lan{\c o}on}},\ and\
  \bibinfo {author} {\bibfnamefont {H.~C.}\ \bibnamefont {Walker}},\ }\bibfield
   {title} {\bibinfo {title} {Evidence for biquadratic exchange in the
  quasi-two-dimensional antiferromagnet {FePS$_3$}},\ }\href
  {https://doi.org/10.1063/5.0009114} {\bibfield  {journal} {\bibinfo
  {journal} {J. Appl. Phys.}\ }\textbf {\bibinfo {volume} {127}},\ \bibinfo
  {pages} {223903} (\bibinfo {year} {2020})}\BibitemShut {NoStop}%
\bibitem [{\citenamefont {Li}\ \emph {et~al.}(2016)\citenamefont {Li},
  \citenamefont {Li}, \citenamefont {Kim}, \citenamefont {Balents},
  \citenamefont {Yu},\ and\ \citenamefont {Chen}}]{li_weyl_2016}%
  \BibitemOpen
  \bibfield  {author} {\bibinfo {author} {\bibfnamefont {F.-Y.}\ \bibnamefont
  {Li}}, \bibinfo {author} {\bibfnamefont {Y.-D.}\ \bibnamefont {Li}}, \bibinfo
  {author} {\bibfnamefont {Y.~B.}\ \bibnamefont {Kim}}, \bibinfo {author}
  {\bibfnamefont {L.}~\bibnamefont {Balents}}, \bibinfo {author} {\bibfnamefont
  {Y.}~\bibnamefont {Yu}},\ and\ \bibinfo {author} {\bibfnamefont
  {G.}~\bibnamefont {Chen}},\ }\bibfield  {title} {\bibinfo {title} {Weyl
  magnons in breathing pyrochlore antiferromagnets},\ }\href
  {https://doi.org/10.1038/ncomms12691} {\bibfield  {journal} {\bibinfo
  {journal} {Nat. Commun.}\ }\textbf {\bibinfo {volume} {7}},\ \bibinfo {pages}
  {12691} (\bibinfo {year} {2016})}\BibitemShut {NoStop}%
\bibitem [{\citenamefont {Hirschberger}\ \emph {et~al.}(2019)\citenamefont
  {Hirschberger}, \citenamefont {Nakajima}, \citenamefont {Gao}, \citenamefont
  {Peng}, \citenamefont {Kikkawa}, \citenamefont {Kurumaji}, \citenamefont
  {Kriener}, \citenamefont {Yamasaki}, \citenamefont {Sagayama}, \citenamefont
  {Nakao}, \citenamefont {Ohishi}, \citenamefont {Kakurai}, \citenamefont
  {Taguchi}, \citenamefont {Yu}, \citenamefont {Arima},\ and\ \citenamefont
  {Tokura}}]{hirschberger_skyrmion_2019}%
  \BibitemOpen
  \bibfield  {author} {\bibinfo {author} {\bibfnamefont {M.}~\bibnamefont
  {Hirschberger}}, \bibinfo {author} {\bibfnamefont {T.}~\bibnamefont
  {Nakajima}}, \bibinfo {author} {\bibfnamefont {S.}~\bibnamefont {Gao}},
  \bibinfo {author} {\bibfnamefont {L.}~\bibnamefont {Peng}}, \bibinfo {author}
  {\bibfnamefont {A.}~\bibnamefont {Kikkawa}}, \bibinfo {author} {\bibfnamefont
  {T.}~\bibnamefont {Kurumaji}}, \bibinfo {author} {\bibfnamefont
  {M.}~\bibnamefont {Kriener}}, \bibinfo {author} {\bibfnamefont
  {Y.}~\bibnamefont {Yamasaki}}, \bibinfo {author} {\bibfnamefont
  {H.}~\bibnamefont {Sagayama}}, \bibinfo {author} {\bibfnamefont
  {H.}~\bibnamefont {Nakao}}, \bibinfo {author} {\bibfnamefont
  {K.}~\bibnamefont {Ohishi}}, \bibinfo {author} {\bibfnamefont
  {K.}~\bibnamefont {Kakurai}}, \bibinfo {author} {\bibfnamefont
  {Y.}~\bibnamefont {Taguchi}}, \bibinfo {author} {\bibfnamefont
  {X.}~\bibnamefont {Yu}}, \bibinfo {author} {\bibfnamefont {T.-h.}\
  \bibnamefont {Arima}},\ and\ \bibinfo {author} {\bibfnamefont
  {Y.}~\bibnamefont {Tokura}},\ }\bibfield  {title} {\bibinfo {title} {Skyrmion
  phase and competing magnetic orders on a breathing kagom\'e lattice},\ }\href
  {https://doi.org/10.1038/s41467-019-13675-4} {\bibfield  {journal} {\bibinfo
  {journal} {Nat. Commun.}\ }\textbf {\bibinfo {volume} {10}},\ \bibinfo
  {pages} {5831} (\bibinfo {year} {2019})}\BibitemShut {NoStop}%
\bibitem [{\citenamefont {Matsumura}\ \emph {et~al.}(2019)\citenamefont
  {Matsumura}, \citenamefont {Ozono}, \citenamefont {Nakamura}, \citenamefont
  {Kabeya},\ and\ \citenamefont {Ochiai}}]{matsumura_helical_2019}%
  \BibitemOpen
  \bibfield  {author} {\bibinfo {author} {\bibfnamefont {T.}~\bibnamefont
  {Matsumura}}, \bibinfo {author} {\bibfnamefont {Y.}~\bibnamefont {Ozono}},
  \bibinfo {author} {\bibfnamefont {S.}~\bibnamefont {Nakamura}}, \bibinfo
  {author} {\bibfnamefont {N.}~\bibnamefont {Kabeya}},\ and\ \bibinfo {author}
  {\bibfnamefont {A.}~\bibnamefont {Ochiai}},\ }\bibfield  {title} {\bibinfo
  {title} {Helical ordering of spin trimers in a distorted kagome lattice of
  {Gd$_3$Ru$_4$Al$_{12}$} studied by resonant x-ray diffraction},\ }\href
  {https://doi.org/10.7566/JPSJ.88.023704} {\bibfield  {journal} {\bibinfo
  {journal} {J. Phys. Soc. Jpn.}\ }\textbf {\bibinfo {volume} {88}},\ \bibinfo
  {pages} {023704} (\bibinfo {year} {2019})}\BibitemShut {NoStop}%
\bibitem [{\citenamefont {Bergman}\ \emph {et~al.}(2007)\citenamefont
  {Bergman}, \citenamefont {Alicea}, \citenamefont {Gull}, \citenamefont
  {Trebst},\ and\ \citenamefont {Balents}}]{bergman_order_2007}%
  \BibitemOpen
  \bibfield  {author} {\bibinfo {author} {\bibfnamefont {D.}~\bibnamefont
  {Bergman}}, \bibinfo {author} {\bibfnamefont {J.}~\bibnamefont {Alicea}},
  \bibinfo {author} {\bibfnamefont {E.}~\bibnamefont {Gull}}, \bibinfo {author}
  {\bibfnamefont {S.}~\bibnamefont {Trebst}},\ and\ \bibinfo {author}
  {\bibfnamefont {L.}~\bibnamefont {Balents}},\ }\bibfield  {title} {\bibinfo
  {title} {Order-by-disorder and spiral spin-liquid in frustrated
  diamond-lattice antiferromagnets},\ }\href {https://doi.org/10.1038/nphys622}
  {\bibfield  {journal} {\bibinfo  {journal} {Nat. Phys.}\ }\textbf {\bibinfo
  {volume} {3}},\ \bibinfo {pages} {487} (\bibinfo {year} {2007})}\BibitemShut
  {NoStop}%
\bibitem [{\citenamefont {Gao}\ \emph {et~al.}(2017)\citenamefont {Gao},
  \citenamefont {Zaharko}, \citenamefont {Tsurkan}, \citenamefont {Su},
  \citenamefont {White}, \citenamefont {Tucker}, \citenamefont {Roessli},
  \citenamefont {Bourdarot}, \citenamefont {Sibille}, \citenamefont
  {Chernyshov}, \citenamefont {Fennell}, \citenamefont {Loidl},\ and\
  \citenamefont {R\"uegg}}]{gao_spiral_2017}%
  \BibitemOpen
  \bibfield  {author} {\bibinfo {author} {\bibfnamefont {S.}~\bibnamefont
  {Gao}}, \bibinfo {author} {\bibfnamefont {O.}~\bibnamefont {Zaharko}},
  \bibinfo {author} {\bibfnamefont {V.}~\bibnamefont {Tsurkan}}, \bibinfo
  {author} {\bibfnamefont {Y.}~\bibnamefont {Su}}, \bibinfo {author}
  {\bibfnamefont {J.~S.}\ \bibnamefont {White}}, \bibinfo {author}
  {\bibfnamefont {G.~S.}\ \bibnamefont {Tucker}}, \bibinfo {author}
  {\bibfnamefont {B.}~\bibnamefont {Roessli}}, \bibinfo {author} {\bibfnamefont
  {F.}~\bibnamefont {Bourdarot}}, \bibinfo {author} {\bibfnamefont
  {R.}~\bibnamefont {Sibille}}, \bibinfo {author} {\bibfnamefont
  {D.}~\bibnamefont {Chernyshov}}, \bibinfo {author} {\bibfnamefont
  {T.}~\bibnamefont {Fennell}}, \bibinfo {author} {\bibfnamefont
  {A.}~\bibnamefont {Loidl}},\ and\ \bibinfo {author} {\bibfnamefont
  {C.}~\bibnamefont {R\"uegg}},\ }\bibfield  {title} {\bibinfo {title} {Spiral
  spin-liquid and the emergence of a vortex-like state in {MnSc$_2$S$_4$}},\
  }\href {https://doi.org/10.1038/nphys3914} {\bibfield  {journal} {\bibinfo
  {journal} {Nat. Phys.}\ }\textbf {\bibinfo {volume} {13}},\ \bibinfo {pages}
  {157} (\bibinfo {year} {2017})}\BibitemShut {NoStop}%
\bibitem [{\citenamefont {Gao}\ \emph {et~al.}(2020)\citenamefont {Gao},
  \citenamefont {Rosales}, \citenamefont {G\'omez~Albarrac\'in}, \citenamefont
  {Tsurkan}, \citenamefont {Kaur}, \citenamefont {Fennell}, \citenamefont
  {Steffens}, \citenamefont {Boehm}, \citenamefont {\v{C}erm\'ak},
  \citenamefont {Schneidewind}, \citenamefont {Ressouche}, \citenamefont
  {Cabra}, \citenamefont {R\"uegg},\ and\ \citenamefont
  {Zaharko}}]{gao_fractional_2020}%
  \BibitemOpen
  \bibfield  {author} {\bibinfo {author} {\bibfnamefont {S.}~\bibnamefont
  {Gao}}, \bibinfo {author} {\bibfnamefont {H.~D.}\ \bibnamefont {Rosales}},
  \bibinfo {author} {\bibfnamefont {F.~A.}\ \bibnamefont
  {G\'omez~Albarrac\'in}}, \bibinfo {author} {\bibfnamefont {V.}~\bibnamefont
  {Tsurkan}}, \bibinfo {author} {\bibfnamefont {G.}~\bibnamefont {Kaur}},
  \bibinfo {author} {\bibfnamefont {T.}~\bibnamefont {Fennell}}, \bibinfo
  {author} {\bibfnamefont {P.}~\bibnamefont {Steffens}}, \bibinfo {author}
  {\bibfnamefont {M.}~\bibnamefont {Boehm}}, \bibinfo {author} {\bibfnamefont
  {P.}~\bibnamefont {\v{C}erm\'ak}}, \bibinfo {author} {\bibfnamefont
  {A.}~\bibnamefont {Schneidewind}}, \bibinfo {author} {\bibfnamefont
  {E.}~\bibnamefont {Ressouche}}, \bibinfo {author} {\bibfnamefont {D.~C.}\
  \bibnamefont {Cabra}}, \bibinfo {author} {\bibfnamefont {C.}~\bibnamefont
  {R\"uegg}},\ and\ \bibinfo {author} {\bibfnamefont {O.}~\bibnamefont
  {Zaharko}},\ }\bibfield  {title} {\bibinfo {title} {Fractional
  antiferromagnetic skyrmion lattice induced by anisotropic couplings},\ }\href
  {https://doi.org/10.1038/s41586-020-2716-8} {\bibfield  {journal} {\bibinfo
  {journal} {Nature}\ }\textbf {\bibinfo {volume} {586}},\ \bibinfo {pages}
  {37} (\bibinfo {year} {2020})}\BibitemShut {NoStop}%
\bibitem [{\citenamefont {Tokura}\ \emph {et~al.}(2017)\citenamefont {Tokura},
  \citenamefont {Kawasaki},\ and\ \citenamefont
  {Nagaosa}}]{tokura_emergent_2017}%
  \BibitemOpen
  \bibfield  {author} {\bibinfo {author} {\bibfnamefont {Y.}~\bibnamefont
  {Tokura}}, \bibinfo {author} {\bibfnamefont {M.}~\bibnamefont {Kawasaki}},\
  and\ \bibinfo {author} {\bibfnamefont {N.}~\bibnamefont {Nagaosa}},\
  }\bibfield  {title} {\bibinfo {title} {Emergent functions of quantum
  materials},\ }\href {https://doi.org/10.1038/nphys4274} {\bibfield  {journal}
  {\bibinfo  {journal} {Nat. Phys.}\ }\textbf {\bibinfo {volume} {13}},\
  \bibinfo {pages} {1056} (\bibinfo {year} {2017})}\BibitemShut {NoStop}%
\end{thebibliography}

\begin{thebibliography}{13}%
\makeatletter
\providecommand \@ifxundefined [1]{%
 \@ifx{#1\undefined}
}%
\providecommand \@ifnum [1]{%
 \ifnum #1\expandafter \@firstoftwo
 \else \expandafter \@secondoftwo
 \fi
}%
\providecommand \@ifx [1]{%
 \ifx #1\expandafter \@firstoftwo
 \else \expandafter \@secondoftwo
 \fi
}%
\providecommand \natexlab [1]{#1}%
\providecommand \enquote  [1]{``#1''}%
\providecommand \bibnamefont  [1]{#1}%
\providecommand \bibfnamefont [1]{#1}%
\providecommand \citenamefont [1]{#1}%
\providecommand \href@noop [0]{\@secondoftwo}%
\providecommand \href [0]{\begingroup \@sanitize@url \@href}%
\providecommand \@href[1]{\@@startlink{#1}\@@href}%
\providecommand \@@href[1]{\endgroup#1\@@endlink}%
\providecommand \@sanitize@url [0]{\catcode `\\12\catcode `\$12\catcode
  `\&12\catcode `\#12\catcode `\^12\catcode `\_12\catcode `\%12\relax}%
\providecommand \@@startlink[1]{}%
\providecommand \@@endlink[0]{}%
\providecommand \url  [0]{\begingroup\@sanitize@url \@url }%
\providecommand \@url [1]{\endgroup\@href {#1}{\urlprefix }}%
\providecommand \urlprefix  [0]{URL }%
\providecommand \Eprint [0]{\href }%
\providecommand \doibase [0]{https://doi.org/}%
\providecommand \selectlanguage [0]{\@gobble}%
\providecommand \bibinfo  [0]{\@secondoftwo}%
\providecommand \bibfield  [0]{\@secondoftwo}%
\providecommand \translation [1]{[#1]}%
\providecommand \BibitemOpen [0]{}%
\providecommand \bibitemStop [0]{}%
\providecommand \bibitemNoStop [0]{.\EOS\space}%
\providecommand \EOS [0]{\spacefactor3000\relax}%
\providecommand \BibitemShut  [1]{\csname bibitem#1\endcsname}%
\let\auto@bib@innerbib\@empty
\bibitem [{\citenamefont {Rodriguez-Carvajal}(1993)}]{rodriguez_recent_1993}%
  \BibitemOpen
  \bibfield  {author} {\bibinfo {author} {\bibfnamefont {J.}~\bibnamefont
  {Rodriguez-Carvajal}},\ }\bibfield  {title} {\bibinfo {title} {Recent
  advances in magnetic structure determination by neutron powder diffraction},\
  }\href {https://doi.org/10.1016/0921-4526(93)90108-I} {\bibfield  {journal}
  {\bibinfo  {journal} {Physica B: Conden. Matter}\ }\textbf {\bibinfo {volume}
  {192}},\ \bibinfo {pages} {55} (\bibinfo {year} {1993})}\BibitemShut
  {NoStop}%
\bibitem [{\citenamefont {Arnold}\ \emph {et~al.}(2014)\citenamefont {Arnold},
  \citenamefont {Bilheux}, \citenamefont {Borreguero}, \citenamefont {Buts},
  \citenamefont {Campbell}, \citenamefont {Chapon}, \citenamefont {Doucet},
  \citenamefont {Draper}, \citenamefont {{Ferraz Leal}}, \citenamefont {Gigg},
  \citenamefont {Lynch}, \citenamefont {Markvardsen}, \citenamefont
  {Mikkelson}, \citenamefont {Mikkelson}, \citenamefont {Miller}, \citenamefont
  {Palmen}, \citenamefont {Parker}, \citenamefont {Passos}, \citenamefont
  {Perring}, \citenamefont {Peterson}, \citenamefont {Ren}, \citenamefont
  {Reuter}, \citenamefont {Savici}, \citenamefont {Taylor}, \citenamefont
  {Taylor}, \citenamefont {Tolchenov}, \citenamefont {Zhou},\ and\
  \citenamefont {Zikovsky}}]{arnold_mantid_2014}%
  \BibitemOpen
  \bibfield  {author} {\bibinfo {author} {\bibfnamefont {O.}~\bibnamefont
  {Arnold}}, \bibinfo {author} {\bibfnamefont {J.}~\bibnamefont {Bilheux}},
  \bibinfo {author} {\bibfnamefont {J.}~\bibnamefont {Borreguero}}, \bibinfo
  {author} {\bibfnamefont {A.}~\bibnamefont {Buts}}, \bibinfo {author}
  {\bibfnamefont {S.}~\bibnamefont {Campbell}}, \bibinfo {author}
  {\bibfnamefont {L.}~\bibnamefont {Chapon}}, \bibinfo {author} {\bibfnamefont
  {M.}~\bibnamefont {Doucet}}, \bibinfo {author} {\bibfnamefont
  {N.}~\bibnamefont {Draper}}, \bibinfo {author} {\bibfnamefont
  {R.}~\bibnamefont {{Ferraz Leal}}}, \bibinfo {author} {\bibfnamefont
  {M.}~\bibnamefont {Gigg}}, \bibinfo {author} {\bibfnamefont {V.}~\bibnamefont
  {Lynch}}, \bibinfo {author} {\bibfnamefont {A.}~\bibnamefont {Markvardsen}},
  \bibinfo {author} {\bibfnamefont {D.}~\bibnamefont {Mikkelson}}, \bibinfo
  {author} {\bibfnamefont {R.}~\bibnamefont {Mikkelson}}, \bibinfo {author}
  {\bibfnamefont {R.}~\bibnamefont {Miller}}, \bibinfo {author} {\bibfnamefont
  {K.}~\bibnamefont {Palmen}}, \bibinfo {author} {\bibfnamefont
  {P.}~\bibnamefont {Parker}}, \bibinfo {author} {\bibfnamefont
  {G.}~\bibnamefont {Passos}}, \bibinfo {author} {\bibfnamefont
  {T.}~\bibnamefont {Perring}}, \bibinfo {author} {\bibfnamefont
  {P.}~\bibnamefont {Peterson}}, \bibinfo {author} {\bibfnamefont
  {S.}~\bibnamefont {Ren}}, \bibinfo {author} {\bibfnamefont {M.}~\bibnamefont
  {Reuter}}, \bibinfo {author} {\bibfnamefont {A.}~\bibnamefont {Savici}},
  \bibinfo {author} {\bibfnamefont {J.}~\bibnamefont {Taylor}}, \bibinfo
  {author} {\bibfnamefont {R.}~\bibnamefont {Taylor}}, \bibinfo {author}
  {\bibfnamefont {R.}~\bibnamefont {Tolchenov}}, \bibinfo {author}
  {\bibfnamefont {W.}~\bibnamefont {Zhou}},\ and\ \bibinfo {author}
  {\bibfnamefont {J.}~\bibnamefont {Zikovsky}},\ }\bibfield  {title} {\bibinfo
  {title} {Mantid—{Data} analysis and visualization package for neutron
  scattering and $\mu${SR} experiments},\ }\href
  {https://doi.org/https://doi.org/10.1016/j.nima.2014.07.029} {\bibfield
  {journal} {\bibinfo  {journal} {Nucl. Instr. Methods Phys. Res. Sec. A}\
  }\textbf {\bibinfo {volume} {764}},\ \bibinfo {pages} {156 } (\bibinfo {year}
  {2014})}\BibitemShut {NoStop}%
\bibitem [{\citenamefont {Toth}\ and\ \citenamefont
  {Lake}(2015)}]{toth_linear_2015s}%
  \BibitemOpen
  \bibfield  {author} {\bibinfo {author} {\bibfnamefont {S.}~\bibnamefont
  {Toth}}\ and\ \bibinfo {author} {\bibfnamefont {B.}~\bibnamefont {Lake}},\
  }\bibfield  {title} {\bibinfo {title} {Linear spin wave theory for single-{Q}
  incommensurate magnetic structures},\ }\href
  {https://doi.org/10.1088/0953-8984/27/16/166002} {\bibfield  {journal}
  {\bibinfo  {journal} {J. Phys.: Condens. Matter}\ }\textbf {\bibinfo {volume}
  {27}},\ \bibinfo {pages} {166002} (\bibinfo {year} {2015})}\BibitemShut
  {NoStop}%
\bibitem [{\citenamefont {Togo}\ and\ \citenamefont
  {Tanaka}(2015)}]{togo_first_2015}%
  \BibitemOpen
  \bibfield  {author} {\bibinfo {author} {\bibfnamefont {A.}~\bibnamefont
  {Togo}}\ and\ \bibinfo {author} {\bibfnamefont {I.}~\bibnamefont {Tanaka}},\
  }\bibfield  {title} {\bibinfo {title} {First principles phonon calculations
  in materials science},\ }\href
  {https://doi.org/https://doi.org/10.1016/j.scriptamat.2015.07.021} {\bibfield
   {journal} {\bibinfo  {journal} {Scr. Mater.}\ }\textbf {\bibinfo {volume}
  {108}},\ \bibinfo {pages} {1 } (\bibinfo {year} {2015})}\BibitemShut
  {NoStop}%
\bibitem [{\citenamefont {Kresse}\ and\ \citenamefont
  {Furthm\"uller}(1996)}]{kresse_efficiency_1996}%
  \BibitemOpen
  \bibfield  {author} {\bibinfo {author} {\bibfnamefont {G.}~\bibnamefont
  {Kresse}}\ and\ \bibinfo {author} {\bibfnamefont {J.}~\bibnamefont
  {Furthm\"uller}},\ }\bibfield  {title} {\bibinfo {title} {Efficiency of
  ab-initio total energy calculations for metals and semiconductors using a
  plane-wave basis set},\ }\href
  {https://doi.org/https://doi.org/10.1016/0927-0256(96)00008-0} {\bibfield
  {journal} {\bibinfo  {journal} {Comput. Mater. Sci.}\ }\textbf {\bibinfo
  {volume} {6}},\ \bibinfo {pages} {15 } (\bibinfo {year} {1996})}\BibitemShut
  {NoStop}%
\bibitem [{\citenamefont {Dudarev}\ \emph {et~al.}(1998)\citenamefont
  {Dudarev}, \citenamefont {Botton}, \citenamefont {Savrasov}, \citenamefont
  {Humphreys},\ and\ \citenamefont {Sutton}}]{dudarev_electron_1998}%
  \BibitemOpen
  \bibfield  {author} {\bibinfo {author} {\bibfnamefont {S.~L.}\ \bibnamefont
  {Dudarev}}, \bibinfo {author} {\bibfnamefont {G.~A.}\ \bibnamefont {Botton}},
  \bibinfo {author} {\bibfnamefont {S.~Y.}\ \bibnamefont {Savrasov}}, \bibinfo
  {author} {\bibfnamefont {C.~J.}\ \bibnamefont {Humphreys}},\ and\ \bibinfo
  {author} {\bibfnamefont {A.~P.}\ \bibnamefont {Sutton}},\ }\bibfield  {title}
  {\bibinfo {title} {Electron-energy-loss spectra and the structural stability
  of nickel oxide: {An LSDA+U} study},\ }\href
  {https://doi.org/10.1103/PhysRevB.57.1505} {\bibfield  {journal} {\bibinfo
  {journal} {Phys. Rev. B}\ }\textbf {\bibinfo {volume} {57}},\ \bibinfo
  {pages} {1505} (\bibinfo {year} {1998})}\BibitemShut {NoStop}%
\bibitem [{\citenamefont {Isseroff}\ and\ \citenamefont
  {Carter}(2012)}]{isseroff_importance_2012}%
  \BibitemOpen
  \bibfield  {author} {\bibinfo {author} {\bibfnamefont {L.~Y.}\ \bibnamefont
  {Isseroff}}\ and\ \bibinfo {author} {\bibfnamefont {E.~A.}\ \bibnamefont
  {Carter}},\ }\bibfield  {title} {\bibinfo {title} {Importance of reference
  hamiltonians containing exact exchange for accurate one-shot {$GW$}
  calculations of {Cu$_2$O}},\ }\href
  {https://doi.org/10.1103/PhysRevB.85.235142} {\bibfield  {journal} {\bibinfo
  {journal} {Phys. Rev. B}\ }\textbf {\bibinfo {volume} {85}},\ \bibinfo
  {pages} {235142} (\bibinfo {year} {2012})}\BibitemShut {NoStop}%
\bibitem [{\citenamefont {Wang}\ \emph {et~al.}(2006)\citenamefont {Wang},
  \citenamefont {Maxisch},\ and\ \citenamefont {Ceder}}]{wang_oxidation_2006}%
  \BibitemOpen
  \bibfield  {author} {\bibinfo {author} {\bibfnamefont {L.}~\bibnamefont
  {Wang}}, \bibinfo {author} {\bibfnamefont {T.}~\bibnamefont {Maxisch}},\ and\
  \bibinfo {author} {\bibfnamefont {G.}~\bibnamefont {Ceder}},\ }\bibfield
  {title} {\bibinfo {title} {Oxidation energies of transition metal oxides
  within the $\mathrm{GGA}+\mathrm{U}$ framework},\ }\href
  {https://doi.org/10.1103/PhysRevB.73.195107} {\bibfield  {journal} {\bibinfo
  {journal} {Phys. Rev. B}\ }\textbf {\bibinfo {volume} {73}},\ \bibinfo
  {pages} {195107} (\bibinfo {year} {2006})}\BibitemShut {NoStop}%
\bibitem [{\citenamefont {Kresse}\ and\ \citenamefont
  {Joubert}(1999)}]{kresse_from_1999}%
  \BibitemOpen
  \bibfield  {author} {\bibinfo {author} {\bibfnamefont {G.}~\bibnamefont
  {Kresse}}\ and\ \bibinfo {author} {\bibfnamefont {D.}~\bibnamefont
  {Joubert}},\ }\bibfield  {title} {\bibinfo {title} {From ultrasoft
  pseudopotentials to the projector augmented-wave method},\ }\href
  {https://doi.org/10.1103/PhysRevB.59.1758} {\bibfield  {journal} {\bibinfo
  {journal} {Phys. Rev. B}\ }\textbf {\bibinfo {volume} {59}},\ \bibinfo
  {pages} {1758} (\bibinfo {year} {1999})}\BibitemShut {NoStop}%
\bibitem [{\citenamefont {Cheng}\ \emph {et~al.}(2019)\citenamefont {Cheng},
  \citenamefont {Daemen}, \citenamefont {Kolesnikov},\ and\ \citenamefont
  {Ramirez-Cuesta}}]{cheng_simulation_2019}%
  \BibitemOpen
  \bibfield  {author} {\bibinfo {author} {\bibfnamefont {Y.~Q.}\ \bibnamefont
  {Cheng}}, \bibinfo {author} {\bibfnamefont {L.~L.}\ \bibnamefont {Daemen}},
  \bibinfo {author} {\bibfnamefont {A.~I.}\ \bibnamefont {Kolesnikov}},\ and\
  \bibinfo {author} {\bibfnamefont {A.~J.}\ \bibnamefont {Ramirez-Cuesta}},\
  }\bibfield  {title} {\bibinfo {title} {Simulation of inelastic neutron
  scattering spectra using {OCLIMAX}},\ }\href
  {https://doi.org/10.1021/acs.jctc.8b01250} {\bibfield  {journal} {\bibinfo
  {journal} {J. Chem. Theory Comput.}\ }\textbf {\bibinfo {volume} {15}},\
  \bibinfo {pages} {1974} (\bibinfo {year} {2019})}\BibitemShut {NoStop}%
\bibitem [{\citenamefont {Haraldsen}\ \emph {et~al.}(2005)\citenamefont
  {Haraldsen}, \citenamefont {Barnes},\ and\ \citenamefont
  {Musfeldt}}]{haraldsen_neutron_2005s}%
  \BibitemOpen
  \bibfield  {author} {\bibinfo {author} {\bibfnamefont {J.~T.}\ \bibnamefont
  {Haraldsen}}, \bibinfo {author} {\bibfnamefont {T.}~\bibnamefont {Barnes}},\
  and\ \bibinfo {author} {\bibfnamefont {J.~L.}\ \bibnamefont {Musfeldt}},\
  }\bibfield  {title} {\bibinfo {title} {Neutron scattering and magnetic
  observables for ${S}=1/2$ spin clusters and molecular magnets},\ }\href
  {http://link.aps.org/doi/10.1103/PhysRevB.71.064403} {\bibfield  {journal}
  {\bibinfo  {journal} {Phys. Rev. B}\ }\textbf {\bibinfo {volume} {71}},\
  \bibinfo {pages} {064403} (\bibinfo {year} {2005})}\BibitemShut {NoStop}%
\bibitem [{\citenamefont {Houchins}\ and\ \citenamefont
  {Haraldsen}(2015)}]{houchins_generalization_2015s}%
  \BibitemOpen
  \bibfield  {author} {\bibinfo {author} {\bibfnamefont {G.}~\bibnamefont
  {Houchins}}\ and\ \bibinfo {author} {\bibfnamefont {J.~T.}\ \bibnamefont
  {Haraldsen}},\ }\bibfield  {title} {\bibinfo {title} {Generalization of
  polarized spin excitations for asymmetric dimeric systems},\ }\href
  {https://doi.org/10.1103/PhysRevB.91.014422} {\bibfield  {journal} {\bibinfo
  {journal} {Phys. Rev. B}\ }\textbf {\bibinfo {volume} {91}},\ \bibinfo
  {pages} {014422} (\bibinfo {year} {2015})}\BibitemShut {NoStop}%
\bibitem [{\citenamefont {Gao}\ \emph {et~al.}(2018)\citenamefont {Gao},
  \citenamefont {Guratinder}, \citenamefont {Stuhr}, \citenamefont {White},
  \citenamefont {Mansson}, \citenamefont {Roessli}, \citenamefont {Fennell},
  \citenamefont {Tsurkan}, \citenamefont {Loidl}, \citenamefont
  {Ciomaga~Hatnean}, \citenamefont {Balakrishnan}, \citenamefont {Raymond},
  \citenamefont {Chapon}, \citenamefont {Garlea}, \citenamefont {Savici},
  \citenamefont {Cervellino}, \citenamefont {Bombardi}, \citenamefont
  {Chernyshov}, \citenamefont {R\"uegg}, \citenamefont {Haraldsen},\ and\
  \citenamefont {Zaharko}}]{gao_manifolds_2018s}%
  \BibitemOpen
  \bibfield  {author} {\bibinfo {author} {\bibfnamefont {S.}~\bibnamefont
  {Gao}}, \bibinfo {author} {\bibfnamefont {K.}~\bibnamefont {Guratinder}},
  \bibinfo {author} {\bibfnamefont {U.}~\bibnamefont {Stuhr}}, \bibinfo
  {author} {\bibfnamefont {J.~S.}\ \bibnamefont {White}}, \bibinfo {author}
  {\bibfnamefont {M.}~\bibnamefont {Mansson}}, \bibinfo {author} {\bibfnamefont
  {B.}~\bibnamefont {Roessli}}, \bibinfo {author} {\bibfnamefont
  {T.}~\bibnamefont {Fennell}}, \bibinfo {author} {\bibfnamefont
  {V.}~\bibnamefont {Tsurkan}}, \bibinfo {author} {\bibfnamefont
  {A.}~\bibnamefont {Loidl}}, \bibinfo {author} {\bibfnamefont
  {M.}~\bibnamefont {Ciomaga~Hatnean}}, \bibinfo {author} {\bibfnamefont
  {G.}~\bibnamefont {Balakrishnan}}, \bibinfo {author} {\bibfnamefont
  {S.}~\bibnamefont {Raymond}}, \bibinfo {author} {\bibfnamefont
  {L.}~\bibnamefont {Chapon}}, \bibinfo {author} {\bibfnamefont {V.~O.}\
  \bibnamefont {Garlea}}, \bibinfo {author} {\bibfnamefont {A.~T.}\
  \bibnamefont {Savici}}, \bibinfo {author} {\bibfnamefont {A.}~\bibnamefont
  {Cervellino}}, \bibinfo {author} {\bibfnamefont {A.}~\bibnamefont
  {Bombardi}}, \bibinfo {author} {\bibfnamefont {D.}~\bibnamefont
  {Chernyshov}}, \bibinfo {author} {\bibfnamefont {C.}~\bibnamefont {R\"uegg}},
  \bibinfo {author} {\bibfnamefont {J.~T.}\ \bibnamefont {Haraldsen}},\ and\
  \bibinfo {author} {\bibfnamefont {O.}~\bibnamefont {Zaharko}},\ }\bibfield
  {title} {\bibinfo {title} {Manifolds of magnetic ordered states and
  excitations in the almost {Heisenberg} pyrochlore antiferromagnet
  {MgCr$_2$O$_4$}},\ }\href {https://doi.org/10.1103/PhysRevB.97.134430}
  {\bibfield  {journal} {\bibinfo  {journal} {Phys. Rev. B}\ }\textbf {\bibinfo
  {volume} {97}},\ \bibinfo {pages} {134430} (\bibinfo {year}
  {2018})}\BibitemShut {NoStop}%
\end{thebibliography}
\end{document}